\documentclass{an}
\usepackage{graphicx,fancyhdr}
\usepackage{times}
\newcommand{\nel}{n_\mathrm{e}}
\newcommand{\nec}{n_\mathrm{c}}
\newcommand{\filfac}{f_\mathrm{v}}

\newcommand{\fdee}{f_\mathrm{D}}
\newcommand{\avnel}{\langle\nel\rangle}
\newcommand{\ovfilfac}{\overline{f}_\mathrm{v}}
\newcommand{\ovnec}{\overline{n}_\mathrm{c}}

\newcommand{\ovfdee}{\overline{f}_\mathrm{D}}
\newcommand{\oviw}{\overline{I}_\mathrm{w}}
\newcommand{\cmcube}{\,{\rm cm^{-3}}}
\newcommand{\cmsix}{\,{\rm cm^{-6}}}

\newcommand{\pc}{\,{\rm pc}}
\newcommand{\kpc}{\,{\rm kpc}}
\newcommand{\EM}{{\rm EM}}

\newcommand{\DM}{{\rm DM}}
\newcommand{\pheins}{\phantom{1}}
\newcommand{\phelf}{\phantom{11}}
\newcommand{\phnote}{\phantom{$^{1)}$}}

\begin{document}

\title{Filling factors and scale heights of the DIG
in the Milky Way}

\author{E.M. Berkhuijsen, D. Mitra, \and P. M\"uller}

\institute{Max-Planck-Institut f\"ur Radioastronomie, Auf dem H\"ugel 69,
           53121 Bonn, Germany}

\date{Received 16 August 2005;
accepted $<$date$>$;
published online $<$date$>$}

\abstract{
The combination of dispersion measures of pulsars, distances from the
model of Cordes \& Lazio (\cite{cordes+lazio02}) and emission measures
from the WHAM survey enabled a statistical study of electron densities
and filling factors of the diffuse ionized gas (DIG) in the Milky Way.
The emission measures were corrected for absorption and contributions
from beyond the pulsar distance. For a sample of 157 pulsars at
$|b|>5\degr$ and $60\degr < \ell < 360\degr$, located in mainly interarm
regions within about 3~kpc from the Sun, we find
that:
{\bf (1)} The average volume filling factor along the line of sight
$\ovfilfac$ and the mean density in ionized clouds $\ovnec$ are
inversely correlated: $\ovfilfac(\ovnec ) = (0.0184\pm 0.0011)
\ovnec^{~-1.07\pm 0.03}$ for the ranges $0.03 < \ovnec < 2\cmcube$ and
$0.8 > \ovfilfac > 0.01$. This relationship is very tight.
The inverse correlation of $\ovfilfac$ and $\ovnec$ causes
the well-known constancy of the average electron density along the
line of sight. As $\ovfilfac(z)$ increases with distance from the
Galactic plane $|z|$, the average size of the ionized clouds increases
with $|z|$.
{\bf (2)} For $|z| < 0.9\kpc$ the local density in clouds $\nec (z)$ and
local filling factor $f(z)$ are inversely correlated because the local
electron density $\nel (z) = f(z) \nec (z)$ is constant. We suggest
that $f(z)$ reaches a maximum value of $>0.3$ near $|z| = 0.9\kpc$,
whereas $\nec (z)$ continues to decrease to higher $|z|$, thus causing
the observed flattening in the distribution of dispersion measures
perpendicular to the Galactic plane above this height.
{\bf (3)} For $|z| < 0.9\kpc$ the local distributions $\nec (z),\ f(z)$
and $\nel^2(z)$ have the same scale height which is in the range $250 <
h \la 500\pc$.
{\bf (4)} The average degree of ionization of the warm atomic gas
$\oviw (z)$ increases towards higher $|z|$ similarly to $\ovfilfac
(z)$. Towards $|z| = 1\kpc$, $\ovfilfac (z) = 0.24\pm 0.05$ and $\oviw
(z) = 0.24\pm 0.02$. Near $|z| = 1\kpc$ most of the warm, atomic
hydrogen is ionized.
}

\correspondence{eberkhuijsen@mpifr-bonn.mpg.de}

\maketitle

%

\section{Introduction}

The interstellar medium (ISM) in galaxies largely consists of hydrogen
that occurs in four different phases: the cold neutral medium (CNM:
molecular gas and cold atomic gas), the warm neutral medium (WNM: warm
atomic gas), the warm ionized medium (WIM) and the hot ionized medium
(HIM). Near the Sun typical temperatures/densities of these phases are
80~K/$40\cmcube$, 8000~K/$0.4\cmcube$, 8000~K/$0.2\cmcube$ and
$10^6$~K/$0.003\cmcube$, respectively (Kulkarni \& Heiles\
\cite{kulkarni+heiles88}; Ferri\`ere\ \cite{ferriere98}), which shows
that they have about equal thermal pressures. Physical processes like
ionization and cooling depend on the structure of the ISM (i.e. diffuse,
or cloudy, clumped), therefore the volume filling factors of the
different phases as well as of the gas within these phases are
important quantities.

Observations of external galaxies and the Milky Way showed that the warm
and hot phases are extended and seem to occupy most of the interstellar
space, whereas the CNM fills only a small fraction (Kulkarni \& Heiles\
\cite{kulkarni+heiles88}; Dickey\ \cite{dickey93}; Ferri\`ere\
\cite{ferriere98}). The WIM consists of dense classical \ion{H}{ii}
regions and diffuse ionized gas (DIG) around and in between these
regions. Also ionized surfaces of cool clouds may contribute to the DIG
(Miller \& Cox\ \cite{miller+cox93}). Walterbos \& Braun
(\cite{walterbos+braun94}) described the wide-spread, diffuse H$\alpha$
emission visible in the spiral arms of M\,31 as tori around the arms
with negligible H$\alpha$ emission in between the arms.

Reynolds (\cite{reynolds91b}) obtained an exponential scale height of
the DIG in the solar neighbourhood of about 900~pc
from the increase of pulsar dispersion measures perpendicular to
the Galactic plane with height above the plane. Scale heights near 1~kpc
were also found by other authors using different pulsar samples
(Bhattacharya \& Verbunt\ \cite{bhatta+verbunt91}; Nordgren et al.\
\cite{nordgren+92}; G\'omez et al.\ \cite{gomez+01}; Cordes \& Lazio\
\cite{cordes+lazio03}). The first estimates
of the volume filling factor of this ionized layer were made by Reynolds
(\cite{reynolds77}), who derived a lower limit to the filling factor of
$\simeq 0.1$ from the emission measures and dispersion measures in the
direction of 24 pulsars at Galactic latitudes $|b| > 5\degr$ and a model
of the ionized gas layer. In a later paper (Reynolds\ \cite{reynolds91a})
he used the dispersion measures and emission measures towards 4 pulsars
in globular clusters about 3~kpc away from the midplane of the disk. He
found a mean filling factor of $\ga 0.2$ which represents an average
value through the full layer. The ionized gas appeared to be
concentrated in extended clouds of mean density $0.08\cmcube$ occupying
about 200~pc along a line of sight towards the Galactic pole.

Maps of neutral and ionized gas of the Milky Way and other edge-on
galaxies show that bright, dense clouds are concentrated to a thin disk
that is surrounded by extended, diffuse emission from less dense gas
reaching large distances from the galaxy plane. This indicates that
there is an anticorrelation between gas density and area filling factor
with the latter increasing away from the plane. In the Milky Way
the volume density of the atomic gas decreases with increasing distance
from the Galactic plane (Dickey \& Lockman\ \cite{dickey+lockman90}),
thus the ionized gas density may also decrease and its volume filling
factor may increase. Kulkarni \& Heiles (\cite{kulkarni+heiles88})
estimated the variation of the local volume filling factor, $f$, and
the local mean density within ionized clouds, $\nec$, with distance
from the plane, $z$, assuming exponential functions for the local
electron densities $\nel$ and $\nel^2$. They obtained $f(z) \simeq 0.11
\exp (|z|/640\pc$) and $\nec (z) = 0.27 \exp (-|z|/360\pc )\cmcube$,
scaled to a Galactic centre distance of 8.5~kpc. The local volume
filling factor then increases from $0.11$ at $z=0\pc$ to $0.52$ at $|z|
= 1\kpc$. At $|z| = 400\pc$, about half the scale height of the DIG, $f
\simeq 0.21$ and $\nec \simeq 0.08\cmcube$, in agreement with Reynold's
(\cite{reynolds91a}) results.

The relationship between the average volume filling factor $\ovfilfac$
and mean electron density within clouds $\ovnec$, both averaged along
the line of sight, carries important information on the structure of the
DIG. For example, Elmegreen (\cite{elmegreen98}, \cite{elmegreen99})
showed that turbulence causing hierarchical, fractal structure leads to
an inverse relationship between $\ovfilfac$ and $\ovnec$.
Pynzar (\cite{pynzar93}) was the first to investigate this relationship,
not only for the DIG but also for the classical \ion{H}{ii} regions. He
used dispersion measures and emission measures towards pulsars and
observations of recombination lines from dense \ion{H}{ii} regions.
These data yielded $\ovfilfac (\ovnec ) \propto \ovnec^{~-0.7}$.
Estimates for a mixture of spiral arm and interarm regions by Heiles et
al. (\cite{heiles+96}) and Berkhuijsen (\cite{elly98}) are in agreement
with Pynzar's result (see compilation by Berkhuijsen\  \cite{elly98}).
An inverse relationship between $\ovfilfac$ and $\ovnec$ is
indicated by the near constancy of the average electron density along
the line of sight $\avnel$ derived from dispersion measures of pulsars
with known distances. The constancy of $\avnel$ was first pointed out by
Weisberg et al. (\cite{weisberg+80}); although it is widely used to
derive pulsar distances, its physical significance did not attract
attention.

Recently, a new masterlist of pulsar data has become available (Hobbs \&
Manchester\ \cite{hobbs+manchester03}). We used the dispersion measures
of pulsars at Galactic latitudes above $|5\degr |$ in this list
together with emission measures from the Wisconsin H$\alpha$ Mapper
(WHAM, Haffner et al.\ \cite{haffner+03}), corrected for absorption and
for contributions from beyond the pulsars, for an improved
determination of the relationship between $\ovfilfac$ and $\ovnec$ in
the DIG. The distances of the pulsars we took from the new model of the
electron distribution in the Galaxy presented by Cordes \& Lazio
(\cite{cordes+lazio02}). In this sample of several hundred pulsars we
also searched for variations of $\ovfilfac$ and $\ovnec$ with $z$ in
the DIG, determined their exponential scale heights and derived the
scale heights of the local functions $f(z)$ and $\nec (z)$.

The paper is organized as follows: In Sect.~2 we describe the
derivation of $\ovfilfac$ and $\ovnec$ and the data used. In Sect.~3 we
explain the statistical procedure and describe the results, which are
discussed in Sect.~4. Section~5 gives a summary of our conclusions.
Appendix~A contains a glossary of the variables used. A table listing
relevant parameters of the pulsars in the final sample is available on
request\footnote{eberkhuijsen@mpifr-bonn.mpg.de}.

\section{Basic relations and data used}

\subsection{Basic relations}
\begin{figure}[htb]
\includegraphics[bb = 48 49 521 166,width=8.5cm,clip=]{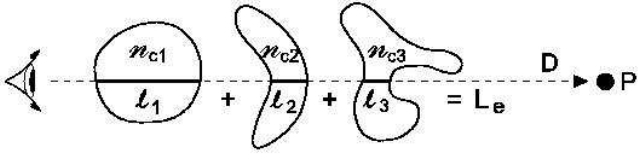}
\caption{Three ionized clouds along the line of sight towards a pulsar
$P$ at distance $D$. The path lengths through the clouds -- $l_1$, $l_2$
and $l_3$ -- are indicated by thick lines. The total path length filled
with electrons is $L_\mathrm{e} = l_1 + l_2 + l_3$ and the filling
fraction $\ovfdee = L_\mathrm{e}/D$. The local electron densities within the
clouds -- $n_\mathrm{c1}$, $n_\mathrm{c2}$ and $n_\mathrm{c3}$ -- give
a mean density in clouds $\ovnec = (n_\mathrm{c1} l_1 + n_\mathrm{c2}
l_2 + n_\mathrm{c3} l_3)/L_\mathrm{e}$. The average electron density
along the line of sight is $\avnel = (n_\mathrm{c1} l_1 + n_\mathrm{c2}
l_2 + n_\mathrm{c3} l_3)/D$. Thus $\avnel = \ovfdee \ovnec$. If
$\ovfdee$ represents the average volume filling factor $\ovfilfac$ (see
Sect.~2.1), $\avnel = \ovfilfac \ovnec$. }
  \label{fig:1}
\end{figure}

The expressions for dispersion measure, DM, and emission measure, EM,
towards a pulsar at a distance $D$ (in pc) can be written in several
ways:

\begin{equation}
 \frac{\DM}{\mathrm{cm}^{-3}\pc} = \int^D_0 \nel (l) dl =
 \avnel D = \ovnec \ovfdee D = \ovnec L_\mathrm{e} \  ,
\label{eq:1}
\end{equation}
\begin{equation}
 \frac{\EM}{\cmsix\pc} = \int^D_0 \nel^2 (l) dl =
 \langle\nel^2\rangle D = \overline{n}^2_\mathrm{c}
 \ovfdee D = \overline{n}^2_\mathrm{c} L_\mathrm{e} \ ,
\label{eq:2}
\end{equation}
where $\nel (l)$ (in $\rm cm^{-3}$) is the electron density at a point
$l$ along the line of sight, $L_\mathrm{e}$ (in pc) the total path
length through the regions containing free electrons (clouds) and
$\ovnec$ (in $\rm cm^{-3}$) the average density in these regions which
is the mean density of a cloud if constant for all clouds along the
line of sight (see Fig.~\ref{fig:1}). Furthermore, $\avnel$ and
$\langle\nel^2\rangle$ are averages along $D$ and $\ovfdee =
L_\mathrm{e}/D$ is the fraction of the line of sight occupied by
electrons. Note that all quantities with overbars are averages
along a line of sight.\footnote{We keep the conventional forms $\avnel$
and $\langle\nel^2\rangle$, but use overbars for other averaged
variables to clearly distinguish similar quantities and to simplify the
notation.}

Combining Eqs. (1) and (2) we find

\begin{equation}
 \ovfdee = \frac{L_\mathrm{e}}{D} = \frac{\avnel}{\ovnec} =
\frac{\langle\nel^2 \rangle}{\overline{n}^2_\mathrm{c}}
\label{eq:3}
\end{equation}
Other useful expressions are:

\begin{equation}
\ovnec = \frac{\DM}{L_\mathrm{e}} = \frac{\EM}{\DM}\, ,\
  L_\mathrm{e} = \frac{\DM^2}{\EM}\ \mathrm{and}\ \ovfdee =
\frac{\DM^2}{\EM\cdot D}\ .
\label{eq:4}
\end{equation}

It is not a priori clear whether the line-of-sight filling factor
$\ovfdee$ is the same as the volume filling factor $\ovfilfac$ that we
want to measure. The volume filling factor is the fraction of the
observed volume occupied by the clouds. Generally, $\ovfilfac = c
\ovfdee$, where $c$ is a factor depending on distance, size and shape
of the clouds. Looking towards a pulsar, the observed volume is the
volume of the radio beam out to the distance of the pulsar. The volume
of this cone is $V_\mathrm{beam} = \frac{\pi}{3} D (D \tan
\frac{\theta}{2})^2$, where $\theta$ is the half-power beamwidth in
degrees. As $\theta$ is small, $\tan \frac{\theta}{2} = \frac{d}{2D}$
and $V_\mathrm{beam} = \frac{\pi}{12} D d^2$, where $d$ is the
linear size of the beam at distance $D$. A thin cloud at distance $D$
and of thickness $l << D$ along the line of sight, more extended than
the beamwidth, occupies $V_\mathrm{c} \simeq \frac{\pi}{12} l d^2$,
so in this case indeed $\ovfilfac \simeq \ovfdee = l/d$ and $c
\simeq 1$. Similarly, $c \simeq 2$ for a spherical cloud of diameter
$d$ equal to the beamwidth, located near the pulsar, and $c\la 1$ for
spherical clouds of diameter $l \la d^{1/3}$. As the DIG mainly
consists of filaments and sheets (Reynolds\ \cite{reynolds91b}), we
could assume $c \simeq 1$ and $\ovfdee = \ovfilfac$. However, an
extended sheet located close to the pulsar will occupy a
larger fraction of the beam volume than a similar cloud near the
observer. On average, clouds will be situated about half-way to the
pulsar, near $D/2$. Then for a cloud thickness $l = \ovfdee D$
the cloud volume is $V_\mathrm{c} = \frac{\pi}{96} D d^2 [(1+
\ovfdee )^3 - (1- \ovfdee )^3]$ and $\ovfilfac
= V_\mathrm{c}/V_\mathrm{beam} = \frac{1}{8} [(1+ \ovfdee )^3
- (1- \ovfdee )^3]$. This gives $c = \ovfilfac /\ovfdee = 0.75$ for
$\ovfdee \la 0.15$ with a slow increase to $c=1$ for larger values of
$\ovfdee$.

In most directions there will be several clouds or sheets along the line
of sight with a total thickness $L_\mathrm{e}$ that is centred near
$D/2$. Therefore we conclude that generally $\ovfdee \simeq
\ovfilfac$ and we will use $\ovfilfac$ in the remaining of the paper.

\subsection{The data}
In order to obtain realistic values for $\ovnec$ and $\ovfilfac$ from
Eqs. (1) to (4) DM and EM should arise in the same ionized regions in
front of the pulsar. For nearby pulsars at low Galactic latitudes a
considerable amount of the observed emission measure EM may originate
beyond the distance of the pulsar, whereas for pulsars at large
distances absorption may reduce EM. Near the Galactic plane classical
\ion{H}{ii} regions on the line of sight towards the pulsar may
also influence the results (Mitra et al.\ \cite{mitra+03}). To minimize
these problems we used pulsars at Galactic latitudes $|b| >
5\degr$.

The dispersion measures of the pulsars were taken from the catalogue of
Hobbs \& Manchester (\cite{hobbs+manchester03}). Among them are 19
pulsars with accurate parallactic distances (Brisken et al.\
\cite{brisken+02}). We checked their positions for bright \ion{H}{ii}
regions, which were found towards 6 of them. We used the 13 distance
calibrators left for comparison with a much larger sample. Of these 13
pulsars 8 are at $|b| > 5\degr$.

The emission measures were taken from the WHAM Northern Sky Survey
(Haffner et al.\ \cite{haffner+03}). This data set represents the total
H$\alpha$ intensity, integrated between LSR velocities $-80$ and
$+80$~km/s, within a one-degree diameter beam centred at a
specified position. The intensities and errors are given in Rayleighs
($\rm 10^6/4\pi \ photons\  cm^{-2}\, s^{-1}\, sr^{-1}$); the
sensitivity of the survey is 0.1R. At the H$\alpha$ wavelength $1
\mathrm{R} = 2.41\, 10^{-7}\ \mathrm{erg}\, \mathrm{cm}^{-2}\,
\mathrm{s}^{-1}\, \mathrm{sr}^{-1}$ which is equivalent to $\EM = 2.25
\cmsix\pc$ for $T_\mathrm{e} = 8000$~K, the mean electron temperature
in the DIG. A possible increase of $T_\mathrm{e}$ with increasing $|z|$
(Haffner et al.\ \cite{haffner+99}) will hardly influence emission
measures observed through the disk at $|b| > 5\degr$.
For the interpolation of these data to a pulsar position we used a
modified quadratic Shepard method for interpolation of scattered data
in a plane and quadratic error propagation. With the constraint $|b| >
5\degr$ the overlap between the pulsar catalogue and WHAM survey yields
a sample of 320 pulsars.

\begin{figure}[htb]
\includegraphics[bb = 69 74 547 699,angle=270,width=8.5cm]{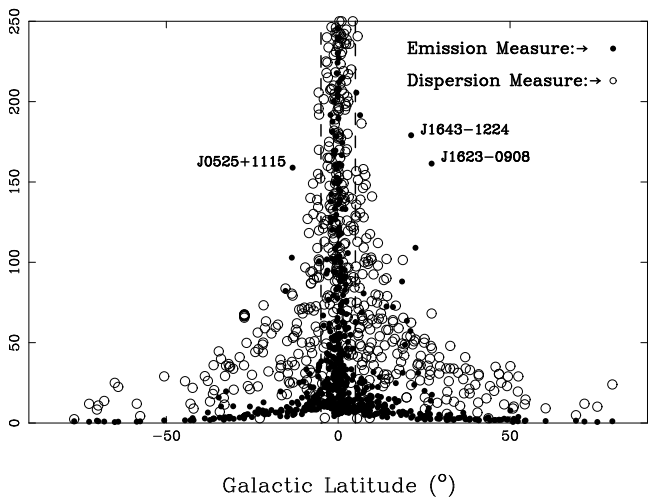}
\caption{Latitude distribution of observed dispersion measures and
emission measures towards 744 pulsars in the field of the WHAM survey
(Haffner et al.\ \cite{haffner+03}), from which the emission measures
were taken. The dashed lines are at $b= \pm 5\degr$. Some pulsars with
exceptionally high emission measures are indicated.}
  \label{fig:2}
\end{figure}

In Fig.~\ref{fig:2} we plot EM and DM as a function of Galactic
latitude. The sharp rise in both EM and DM at $|b| < 5\degr$ is due to
the classical \ion{H}{ii} regions and increasing lines of sight. In
most cases $\EM < \DM$, which indicates that generally $\ovnec <
1\cmcube$. For all further analyses we excluded 30 more pulsars from
our $|b| > 5\degr$ sample based on the fact that they deviate
significantly from the outer envelope of the EM distribution. We also
excluded pulsars located in the Magellanic clouds and in globular
clusters. Our sample at $|b| > 5\degr$ then consisted of 285 pulsars
including 8 of the 13 distance calibrators. The relevant parameters of
the 13 calibrators are listed in Table~\ref{tab:1}.

We determined the distances towards the pulsars in our sample from the
new model of the electron distribution in the Galaxy of Cordes
\& Lazio (\cite{cordes+lazio02}) which is the most realistic model
presently available. Although its basic structure (i.e. thick disk,
thin disk, spiral arms) is nearly identical to that of the older models
of Taylor \& Cordes (\cite{taylor+cordes93}) and G\'omez et al.
(\cite{gomez+01}), it contains more detail resulting from the larger
data sets of various kinds on which it is based (i.e. not only on
dispersion measures, but also on scattering measures as well as optical
and X-ray observations). Therefore this model performs considerably
better than the earlier models (see Table~1 in Cordes \& Lazio
(\cite{cordes+lazio03}). It is well defined for the pulsars in our final
sample, which are at $|b|> 5\degr$ and within about 3~kpc from the Sun
in mainly interarm regions (see Fig.~\ref{fig:3}). The statistical
error in the model distances is about 20\% (Cordes \& Lazio\
\cite{cordes+lazio02}, their Fig.~12), which is much smaller than the
intrinsic spread in the observed dispersion measures (s.d. of factor 2)
and emission measures (s.d. of factor 4, see Fig.~\ref{fig:5}).
This means that the errors in the model distances are {\em negligible}
in our statistical study of the relationship between the electron
density in clouds and filling factors in the DIG, and of their
variation with height above the Galactic plane.

\begin{table*}[htb]
\caption{Parameters of 13 pulsars with parallactic distances }
\label{tab:1}
\begin{tabular}{lrrcccr@{$\pm$}lccc}
\hline
\hline
\noalign{\medskip}
\multicolumn{1}{c}{PSR}&
\multicolumn{1}{c}{$\ell$}&
\multicolumn{1}{c}{$b$}&
\multicolumn{1}{c}{DM$^{1)}$}&
\multicolumn{1}{c}{$D$$^{2)}$}&
\multicolumn{1}{c}{$|z_\mathrm{p}|$}&
\multicolumn{2}{c}{EM}&
\multicolumn{1}{c}{$\EM_\mathrm{p}$$^{3)}$}&
\multicolumn{1}{c}{$\ovnec$}&
\multicolumn{1}{c}{$\ovfilfac$}\\
\multicolumn{1}{c}{Jname}&
\multicolumn{1}{c}{[$\degr$]}&
\multicolumn{1}{c}{[$\degr$]}&
\multicolumn{1}{c}{[$\cmcube\pc$]}&
\multicolumn{1}{c}{[kpc]}&
\multicolumn{1}{c}{[kpc]}&
\multicolumn{2}{c}{[$\cmsix\pc$]}&
\multicolumn{1}{c}{[$\cmsix\pc$]}&
\multicolumn{1}{c}{[$\cmcube$]}&
\multicolumn{1}{c}{}\\
\noalign{\medskip}
\hline
\noalign{\medskip}

J0332+5434 &145.0 &$-$1.2 &26.78 &$1.03\pheins ^{+0.13}_{-0.12}$\pheins
    &0.022$\pm$0.003  &10.0&2.9  &46$\pm$18 &1.7\pheins $\pm$0.7\pheins
    &0.015$\pm$0.006 \\
J0814+7429 &140.0 &31.6 &\pheins 5.75  &0.433$\pm$0.008 &0.227$\pm$0.004
    &1.79&0.03  &\pheins 1.2$\pm$0.1  &0.20$\pm$0.02 &0.065$\pm$0.005 \\
J0953+0755 &228.9 &43.7 &\pheins 2.97  &0.262$\pm$0.005 &0.181$\pm$0.003
    &3.6&0.9  &\pheins 1.9$\pm$0.5  &0.65$\pm$0.17 &0.017$\pm$0.005 \\
J1136+1551 &241.9 &69.2 &\pheins 4.85  &0.35\pheins $\pm$0.02
    &0.327$\pm$0.019 &1.5&0.4 &\pheins 1.1$\pm$0.3 &0.23$\pm$0.06
    &0.059$\pm$0.017\\
J1932+1059 &47.4  &$-$3.9 &\pheins 3.18  &0.331$\pm$0.010
    &0.023$\pm$0.001 &10.5&1.1 &\pheins 2.9$\pm$0.5 &0.92$\pm$0.16
    &0.010$\pm$0.002\\
J2018+2839 &68.1  &$-$4.0 &14.18 &0.95\pheins$\pm$0.09\pheins
    &0.066$\pm$0.006 &13.2&0.2  &\pheins 9.5$\pm$1.4 &0.67$\pm$0.10
    &0.022$\pm$0.004 \\
J2022+2854 &68.9  &$-$4.7 &24.62 &2.3\phelf$^{+1.0}_{-0.6}$
    &0.19\pheins$^{+0.08}_{-0.05}$ &18.2&1.2 &25.4$\pm$3.1
    &1.03$\pm$0.13  &0.010$^{+0.005}_{-0.003}$\\
J2022+5154 &87.9  &8.4  &22.58 &1.9\phelf$^{+0.3}_{-0.2}$
    &0.28\pheins$^{+0.04}_{-0.03}$  &10.5&2.1 &11.9$\pm$2.5
    &0.53$\pm$0.11  &0.023$^{+0.006}_{-0.005}$\\
J0922+0638 &225.4 &36.4 &27.31 &1.15\pheins$^{-0.21}_{-0.16}$
    &0.68\pheins$^{+0.12}_{-0.09}$ &\pheins 6.4&1.1
    &\pheins 6.7$\pm$1.2  &0.24$\pm$0.04 &0.097$^{+0.025}_{-0.022}$\\
J1537+1155 &19.9  &48.3 &11.62 &1.08\pheins$^{+0.16}_{-0.14}$
    &0.81\pheins$^{+0.12}_{-0.10}$ &\pheins 2.7&0.2
    &\pheins 2.9$\pm$0.2 &0.25$\pm$0.02 &0.044$^{+0.007}_{-0.006}$\\
J1857+0943 &42.3  &3.1  &13.31 &0.79\pheins$^{+0.29}_{-0.17}$
    &0.043$^{+0.015}_{-0.009}$ &18.6&2.9 &13.0$^{+5.0}_{-3.7}$
    &0.98$^{+0.37}_{-0.28}$ &0.017$^{+0.009}_{-0.006}$\\
J1713+0747 &28.8  &25.2 &15.99 &0.90\pheins$^{+0.04}_{-0.02}$
    &0.383$^{+0.017}_{-0.009}$ &\pheins 4.3&0.2 &\pheins 3.9$\pm$0.3
    &0.25$\pm$0.02 &0.072$\pm$0.006 \\
J1744$-$1134 &14.8  &9.2  &\pheins 3.14  &0.35\pheins$^{+0.03}_{-0.02}$
    &0.056$^{+0.005}_{-0.003}$ &12.0&0.3 &\pheins 3.7$^{+0.5}_{-0.4}$
    &1.19$^{+0.16}_{-0.14}$ &0.008$\pm$0.001 \\
\noalign{\smallskip}
\hline
\noalign{\medskip}
\multicolumn{11}{l}{$^{1)}$ Taylor et al. (\cite{taylor+93}), errors are
$\la 1\%$; $^{2)}$ Brisken et al. (\cite{brisken+02})}\\
\multicolumn{11}{l}{$^{3)}$ Emission measure in front of the pulsar,
corrected for absorption (see Sect. 2.3). Errors are one standard deviation}\\
\end{tabular}
\end{table*}

\subsection{Corrections to emission measures}
The emission measures observed in the directions of the pulsars
are not the ideal values for obtaining filling factors and electron
densities from Eqs.~(1) to (4). On the one hand they are too small
because of absorption of the H$\alpha$ emission by dust along the line
of sight, while on the other hand they are too large because of
contributions from distances beyond the pulsars. We have developed
approximate corrections for these two competing effects.

First we correct the observed emission measures for absorption.
Following Haffner et al. (\cite{haffner+98}), the corrected emission
measure is
$$ \EM_\mathrm{c} = \EM \cdot A\quad \mathrm{with}\ A = e^{\rm
2.2E(B-V)} \ , $$
where $\rm E(B-V)$ is the colour excess along the line of sight. We can
estimate $\rm E(B-V)$ from the results of Diplas \& Savage
(\cite{diplas+savage94}) who obtained $\mathrm{E(B-V)} \sin |b|$ as a
function of distance $|z|$ above the Galactic plane for a large number
of stars. As they selected stars seen through the diffuse ISM, avoiding
dust clouds, their result is especially suited for correcting the
emission measures from the DIG. They derived an exponential scale
height of the dust layer of $152\pm 7\pc$ and $[\mathrm{E(B-V)}/
\mathrm{kpc}]= 0.257\pm .010\, \mathrm{mag\, kpc}^{-1}$ in the
Galactic plane. Integration of this function to infinite $|z|$ yields
$\mathrm{E(B-V)} \sin|b| = (0.0391\pm 0.0024)$ and $A = \exp ((0.086\pm
0.005)/\sin |b|)$. As our sample only contains pulsars at $|b| >
5\degr$, the largest value of $A=2.68\pm 0.16$ (see Fig.~\ref{fig:4})
and we expect that the
observed and corrected emission measures contain contributions from all
distances along the line of sight through the electron layer. The data
of Diplas \& Savage (\cite{diplas+savage94}) show considerable spread
around their fitted line which they attribute to inhomogeneities in the
dust distribution. These will cause variations in the absorption of the
H$\alpha$ emission and extra spread in the observed emission
measures.\footnote{The function of $A$ derived from the work of
Diplas \& Savage (\cite{diplas+savage94}) at $|b| \ga 10\degr$ is
consistent with Eq.~1 given by Dickinson et al. (\cite{dickinson+03})
which is based on the dust map of Schlegel et al. (\cite{schlegel+98}).
At lower latitudes our values are smaller than those of Dickinson et
al., because we only consider diffuse dust. }

\begin{figure}[htb]
\includegraphics[bb = 136 100 524 397,angle=270,width=8.4cm]{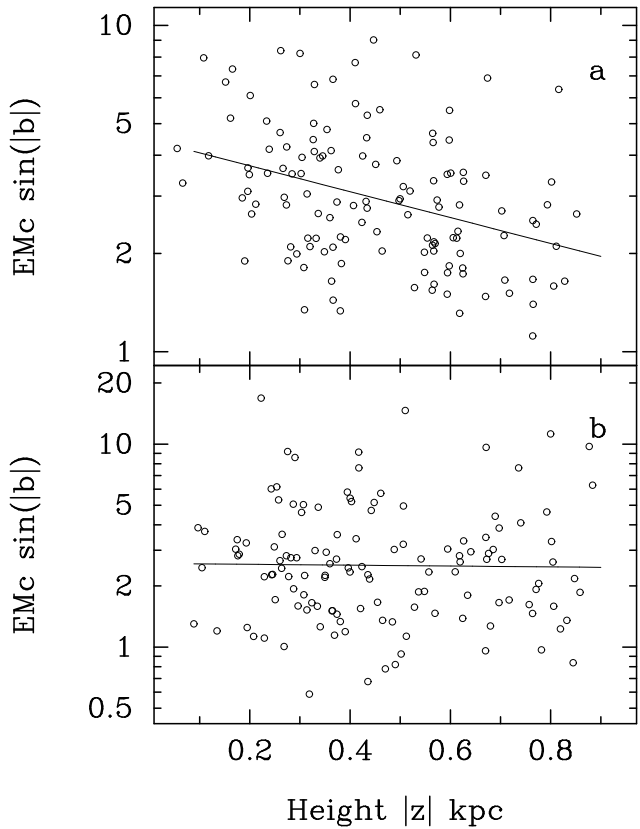}
\caption{Dependence of the emission measures (in $\rm cm^{-6}\, pc$)
perpendicular to the Galactic plane, corrected for absorption, on
distance to the plane. The full lines are least-squares regression
lines of $\log (\EM_\mathrm{c} \sin|b|)$ on $|z|$. If the absorption
corrections are adequate $\EM_\mathrm{c} \sin|b|$ should be independent
of $|z|$ as is the case for pulsars at $60\degr < \ell < 360\degr$
({\bf b}), but not for pulsars at $0\degr < \ell < 60\degr$ ({\bf a})
where the slope of the line is significant
$(-0.40\pm0.09)$. }
  \label{fig:new3}
\end{figure}

\begin{figure}[htb]
\includegraphics[bb = 136 21 487 759,angle=270,width=8.5cm]{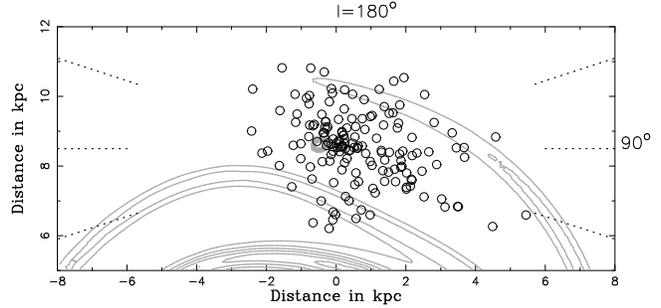}
\caption{Distribution of the final sample of 157 pulsars projected on
the Galactic plane. Full lines are contours of electron density from the
model of Cordes \& Lazio (\cite{cordes+lazio02}). Dotted lines
indicate Galactic longitude. The Sun is located at 8.5~kpc from the
Galactic centre. }
  \label{fig:3}
\end{figure}

\begin{figure}[htb]
\includegraphics[bb = 92 31 578 703,angle=270,width=8.5cm]{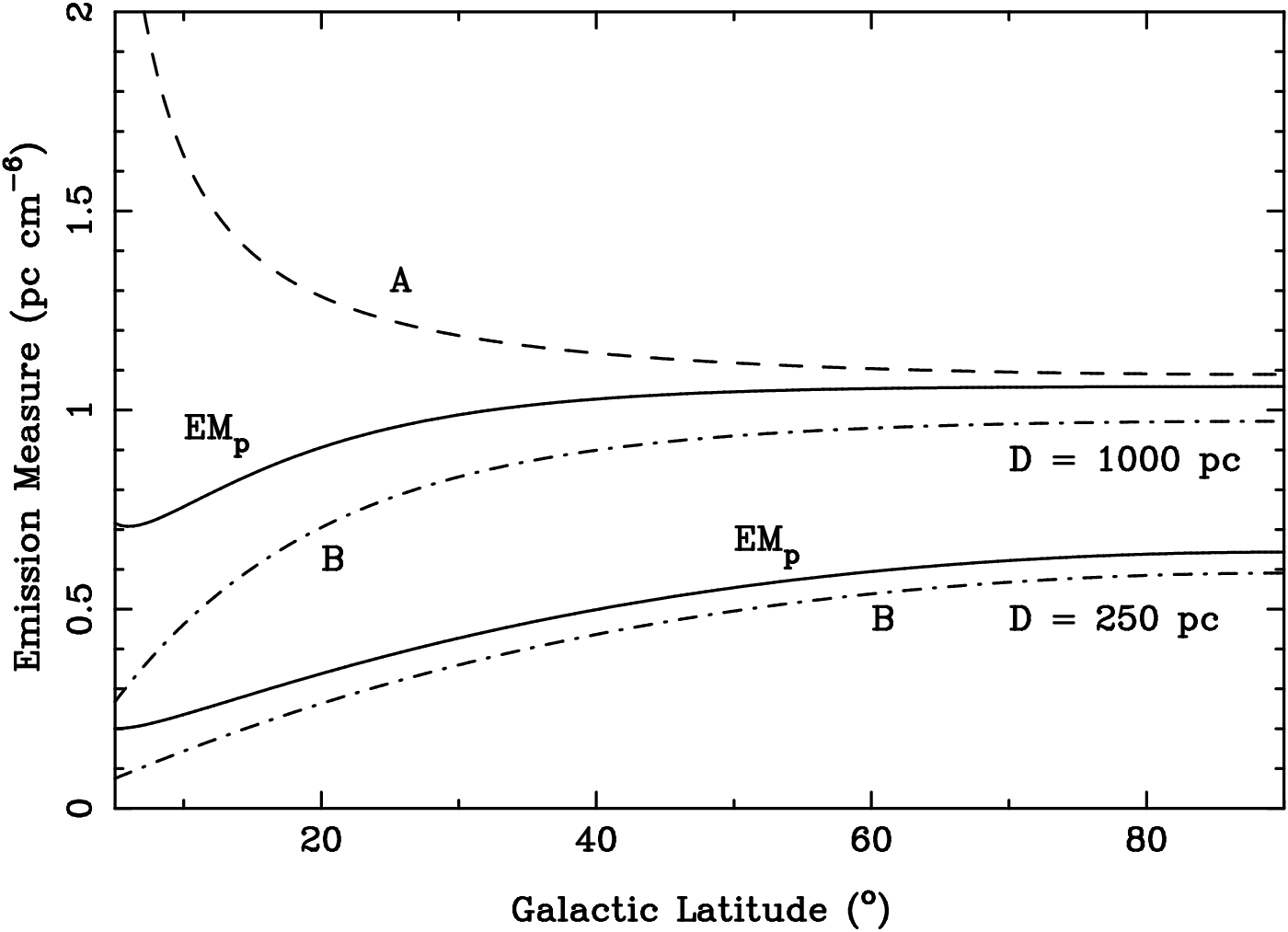}
\caption{Examples of the corrections to the observed emission measure as
a function of Galactic latitude for an assumed value $\EM = 1\cmsix\pc$
towards pulsars at distances $D=250$~pc and 1~kpc. Upper dashed line:
correction for absorption, $A= e^{0.086/\sin|b|}$.
The standard deviation in $A$ decreases from 0.16 at $|b| = 5\degr$
to 0.05 at $|b| = 10\degr$, 0.02 at $|b| = 20\degr$ and $<0.01$ at $|b|
> 40\degr$. Dot-dashed lines: correction for contributions to EM from
beyond the pulsar distance, $B=1-e^{-|z_\mathrm{p}|/h}$ with
$h=280$~pc. For $h = 500$~pc $B$ is 0.1 to 0.2 lower than the curves
shown. The standard deviation in $B$ is $< 0.08$ for $250 < h <
500$~pc. Full lines: corrected emission measure $\EM_\mathrm{p} = \EM
\cdot A \cdot B$. The errors in $\EM_\mathrm{p}$ are typically
$\sim15$\%; they are dominated by the standard deviations in the
observed EM. See Sect.~2.3 for further explanations.
}
  \label{fig:4}
\end{figure}

Since we expect that $\EM_\mathrm{c}$ contains contributions from the
full line of sight through the electron layer, $\EM_\mathrm{c} \sin|b|$
for a pulsar represents the absorption-free emission measure of the full
layer perpendicular to the Galactic plane at the distance of the
pulsar. It will vary among pulsars due to spread in the data and true
variations in electron density with position. But if the absorption
correction is applicable, the values of $\EM_\mathrm{c} \sin|b|$ for
the pulsar sample should be independent of $|z|$. We checked this for
our total sample and found no dependence on $|z|$ for longitudes $\ell
> 60\degr$, but a significant decrease of $\EM_\mathrm{c} \sin|b|$ with
$|z|$ for $0\degr < \ell < 60\degr$ (see Fig.~\ref{fig:new3}).
This could be due to unusual absorption along these lines of sight.
The area contains an extended dust cloud with high $\mathrm{E(B-V)}/
\mathrm{kpc}$ at 300~pc distance from the Sun (Sherwood\
\cite{sherwood74}; Lucke\ \cite{lucke78}) and a chain of dense molecular
clouds known as the Aquila Rift (Dame et al.\ \cite{dame+01}) extending
to $b > +10\degr$. As more than 70\% of the pulsars at $0\degr < \ell <
60\degr$ are at $|b|< 15\degr$, many lines of sight will cut through
dust clouds. In a plot of $\EM$ against longitude the effect of the dust
clouds is clearly visible as a deep minimum centred on $\ell = 40\degr$
which does not occur in the dispersion measures. Such disagreement is
not seen at $\ell > 60\degr$. Because the absorption correction taken
from Diplas \& Savage (\cite{diplas+savage94}) only applies to the
diffuse dust, we had to omit 128 pulsars in the range $0\degr < \ell <
60\degr ,\  b > -15\degr$ from our sample, leaving 157 pulsars at $|b|
> 5\degr$ for further analysis. Of these, 14 pulsars are at $0\degr <
\ell < 60\degr,\  b < -15\degr$. Only 5 distance calibrators of
Table~\ref{tab:1} are left in the sample. The distribution of these 157
pulsars projected on the Galactic plane is shown in Fig.~\ref{fig:3}.
Most of the pulsars in our final sample are located in interarm regions
and within about 3~kpc from the Sun.

Next we determine which part of the corrected emission measure
originates in front of the pulsar. The absorption-corrected emission
measure increases with distance from the Galactic plane as

\begin{equation}
\EM_\mathrm{c}(z) \sin |b| = \int_0^\mathrm{|z|} \nel^2 (z) dz =
\int_0^\mathrm{|z|} \nel^2(0)\, e^\mathrm{-|z|/h}  dz \  ,
\label{eq:5}
\end{equation}
where $\nel^2(0)$ is the value of $\nel^2(z)$ at $|z|=0\pc$ and $h$ is
the exponential scale height. The fraction of the total emission measure
towards the pole coming from the layer below the $|z|$-distance of the
pulsar, $|z_\mathrm{p}|$, is the ratio of Eq.~(5) integrated to
$|z_\mathrm{p}|$ and that integrated through the full layer. Hence, in
front of the pulsar originates

\begin{equation}
\EM_\mathrm{p} \sin |b| = \EM_\mathrm{c} \sin |b| \cdot (1 -
  e^\mathrm{-|z_p|/h})  \ .
\end{equation}

At this point we do not know the scale height $h$, but we do know that
the mean value of $\EM_\mathrm{c} \sin |b|$ for all pulsars should equal
$\nel^2(0) h$ for the sample considered. In Sect.~4.1 we combine
$\overline{\EM_\mathrm{c} \sin|b|}$ with information on $\DM\sin|b|(z)$
to evaluate $h$, which appears to be in the range $250 < h < 500\pc$.
The minimum value of $280\pm 30\pc$ is best constrained. Therefore we
used $h = 280\pc$ in Eq.~(6). A larger value of $h$ leads to a smaller
value of the correction factor $B = 1 - e^\mathrm{-|z_p|/h}$.
We show in Sect.~3.5 that our results are not very sensitive to the
choice of $h$ in the range $250 < h < 500\pc$.

In Fig.~\ref{fig:4} we illustrate the effect of the two corrections to
$\EM$ for pulsars at different latitudes and distances. Especially for
pulsars at low $|z|$ (i.e. small $D$, low $b$) the correction for the
contribution to $\EM$ from beyond the pulsar dominates the absorption
correction leading to $\EM_\mathrm{p} < \EM$.

\begin{figure*}[htb]
\includegraphics[bb = 48 271 528 516,width=16cm,clip=]{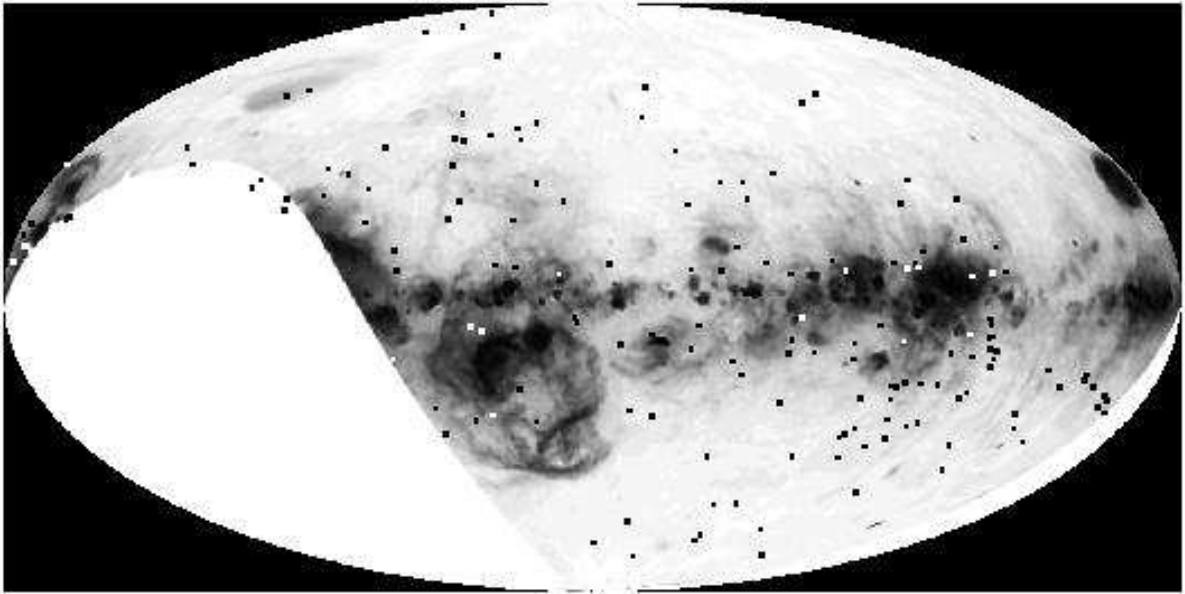}
\caption{Positions of the 157 pulsars in the final sample (filled
squares) projected onto the distribution of the H$\alpha$ emission (WHAM
survey, Haffner et al.\ \cite{haffner+03}). The map is centred at $\ell
= 180\degr,\ b= 0\degr$ and $\ell = 0\degr$ is at the right-hand border.
}
  \label{fig:new6}
\end{figure*}

The final sample of 157 pulsars projected on the WHAM survey is shown in
Fig.~\ref{fig:new6}. A table collecting the properties of these pulsars
is available on request\footnote{eberkhuijsen@mpifr-bonn.mpg.de}.

\section{Results}
In this section we compare the observed as well as the corrected EM
with the DM towards the 157 pulsars in our sample (Sect.~3.1), derive
the exponential scale heights of $\avnel$,
$\langle\nel^2\rangle$, $\ovnec$ and $\ovfilfac$ (Sect.~3.2), and
present the relationship between $\ovfilfac$ and $\ovnec$ (Sect.~3.3).

As EM and DM do not directly depend on each other, a
symmetrical--statistical treatment is needed to find their relationship.
Therefore we calculated the ordinary least-squares bisector in the
log--log plane (Isobe et al.\ \cite{isobe+90}) using the modified
program code distributed by Feigelson. In all other cases we derived the
ordinary least-squares regression line of $Y$ on $X$ in the relevant
units. The quality of a correlation can be judged from the correlation
coefficient $r$ and the value of $t$ resulting from the Student
$t$-test, which gives the probability that the correlation is
accidental. The correlation is highly significant at the $3\sigma$ level
if $t> 3.1$ for a number of independent points $N>100$. The values of
$r$, the standard deviation in $r,\ \sigma_\mathrm{r}$, and $t$ are
calculated from the equations

\begin{displaymath}
r = \frac{\Sigma_1^\mathrm{N} (X_i - \overline{X}) (Y_i -
\overline{Y})}{N \sigma_\mathrm{X} \sigma_\mathrm{Y}}\ ,
\quad
\sigma_\mathrm{r} = \sqrt{\frac{1-r^2}{N-2}}\ ,
\end{displaymath}

\begin{displaymath}
t = r \sqrt{\frac{N-2}{1-r^2}} \ ,
\end{displaymath}
where $\overline{X}$ and $\overline{Y}$ are the mean values of $X$ and
$Y$, respectively, and $\sigma_\mathrm{X}$ and $\sigma_\mathrm{Y}$ the
standard deviations in $X$ and $Y$. The measurement errors in EM and
DM are small compared to the uncertainty of $\la20\%$ in $D$ (Cordes \&
Lazio\ \cite{cordes+lazio02}). Since all these errors are much smaller
than the intrinsic scatter of the points ($\ga$ factor 2), all points
were given equal weight in the fitting procedures. The results are
listed in Table~\ref{tab:2}.

\subsection{Dispersion measures and emission measures}
\begin{figure}[htb]
\includegraphics[bb = 80 21 533 764,angle=270,width=8.5cm]{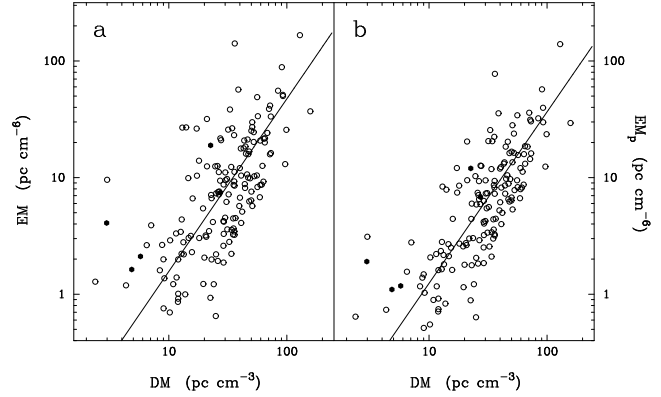}
\caption{Comparison of emission measure and dispersion measure for the
sample of 157 pulsars. {\bf (a)} Observed emission measures, EM.
{\bf (b)} Emission measures corrected for absorption and
contributions from beyond the pulsar distance, $\EM_\mathrm{p}$. The
full lines indicate the bisector fits given in Table~\ref{tab:2}. Filled
circles are pulsars with parallactic distances. }
\label{fig:5}
\end{figure}

If EM and DM occur in the same ionized regions along the line of sight,
we expect $\EM = \ovnec \DM$, or $\EM \propto \DM^\mathrm{b}$
with $b>1$. To test this we plot in Fig.~\ref{fig:5}a the uncorrected
emission measures EM as a function of DM and in Fig.~\ref{fig:5}b the
corrected emission measures $\EM_\mathrm{p}$ against DM for all 157
pulsars at $|b| > 5\degr$. The corrections made to EM have clearly
reduced the spread in the data. The bisector fit in Fig.~\ref{fig:5}b
yields $\EM_\mathrm{p} \propto \DM^{1.47\pm 0.09}$ and with $r=0.76$
and $t=15$ the correlation is highly significant. Because of the even
distribution in latitude of the pulsars and the large fraction of
pulsars (72\%) at $D > 1$~kpc, the slope of the bisector has not
changed after the corrections to EM (see Table~\ref{tab:2}). Note that
the components parallel to the Galactic plane, $\EM_\mathrm{p} \cos|b|$
and $\DM \cos |b|$, are even better correlated with nearly the same
exponent.

Pynzar (\cite{pynzar93}) obtained $\EM \propto \DM^2$ from the lower
envelope to the data in his sample. This much steeper slope could be
due to an increasing overestimate of the emission measures towards
pulsars at large distances. Overestimates of $\EM$ could arise from the
properties of the spiral arm model used by him or from contributions to
$\EM$ from \ion{H}{ii} regions along the line of sight, especially as
two thirds of the pulsars he used are at latitudes below $|5\degr |$.

\begin{figure}[htb]
\includegraphics[bb = 136 111 533 677,angle=270,width=8.5cm]{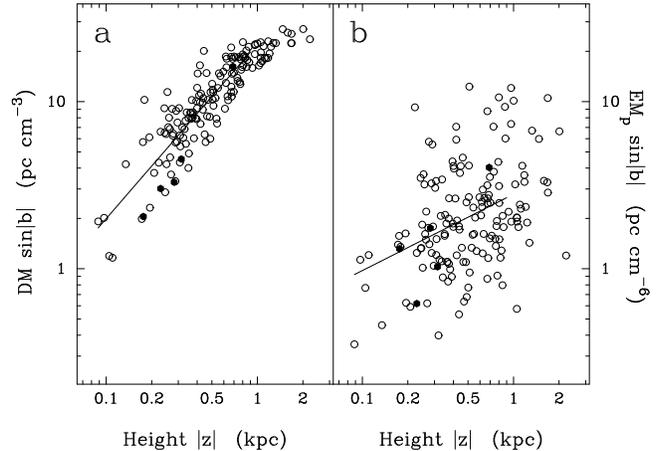}
\caption{{\bf (a)} Dependence of the dispersion measure
perpenciular to the Galactic plane, $\DM\sin|b|$, on distance above the
plane, $|z|$. {\bf (b)} Same for the corrected emission measure
perpendicular to the plane, $\EM_\mathrm{p}\sin|b|$. Full lines indicate
the power-law fits given in Table~\ref{tab:2}. Filled circles are
pulsars with parallactic distances. }
  \label{fig:6}
\end{figure}

The variations of $\DM \sin|b|$ and $\EM_\mathrm{p}\sin|b|$ with height
above the Galactic plane, $|z|$, depend on the vertical distribution of
the ionized gas. Figure~\ref{fig:6}a shows that $\DM\sin|b|$
continuously increases up to $|z| \simeq 0.9\kpc$ and then levels off.
This behaviour reflects that of the more than 100 pulsars with
independent distance estimates on which the model of Cordes \& Lazio
(\cite{cordes+lazio02}) is partly based. A power-law fit to the data
in Fig.~\ref{fig:6}a for $|z|<0.9\kpc$ yields $\DM \sin |b|= (22.1\pm
1.3) |z|_\mathrm{kpc}^{1.04\pm 0.05} \cmcube\pc$ and is highly
significant (see Table~\ref{tab:2}).

Figure~\ref{fig:6}b shows that $\EM_\mathrm{p} \sin|b|$ also increases
with $|z|$ although not as strongly as $\DM\sin|b|$. The power-law fit
for $|z|<0.9\kpc$ gives $\EM_\mathrm{p}\sin|b| = (2.78\pm 0.28)\
|z|_\mathrm{kpc}^{0.46\pm 0.11}$ $\!\cmsix\pc$ with high significance.
It is not clear whether $\EM_\mathrm{p}\sin|b|$ increases beyond
$|z|>1\kpc$, because of the large spread. The value of
$\overline{\EM_\mathrm{c}\sin|b|}$ for the 130 pulsars at $|z| < 0.9\kpc$
is $3.18\pm 0.19\cmsix\pc$. Without absorption correction the mean
value of $\EM\sin|b|$ is $2.18\pm 0.13\cmsix\pc$ \footnote{This
value, corresponding to $0.97\pm 0.06R$, is consistent with the mean
H$\alpha$ intensity of the survey of Haffner et al. (\cite{haffner+03},
their Fig. 13) at $|b| = 25\degr$, the mean latitude of the 130 pulsars
in our sample. }, thus the mean absorption towards the poles is a
factor 1.5.

Comparison of Figs.~\ref{fig:6}a and \ref{fig:6}b reveals the much
larger spread in $\EM_\mathrm{p}$ than in DM. Part of this extra spread
arises because fluctuations in electron density influence DM linearly
but $\EM_\mathrm{p}$ quadratically. The clumpy structure of the
dust (Diplas \& Savage\ \cite{diplas+savage94}) leads to variations in
absorption and increases the spread in the observed emission measures.
Inadequate corrections for absorption and for contributions from beyond
the pulsars also enhance the spread in $\EM_\mathrm{p}$ compared to
that in $\DM$.

\subsection{Scale heights of $\avnel$, $\langle \nel^2\rangle$,
$\ovnec$ and $\overline{f}_\mathrm{v}$ }

The variations of $\ovnec$ and $\ovfilfac$ with height above the
Galactic plane are important functions for the understanding
of physical processes in the warm ionized disk. Therefore we
investigate the dependencies of $\avnel$, $\langle\nel^2\rangle$,
$\ovnec$ and $\ovfilfac$ on $|z|$ in this section. Although these
variables represent averages along the line of
sight\footnote{Averages along the line of sight are equal to averages
along $|z|$. For example, $\avnel = \DM/D = \DM\sin|b|/|z|$.}, we
approximate their $|z|$-distributions by exponentials and estimate the
scale heights $H$. The relationship between $H$ and the scale height
$h$ of the local electron density $\nel^2 (z)$ is given by Eq.~(9) in
Sect.~3.5. Because $\DM \sin |b|(z)$ flattens near $|z| \simeq
1\kpc$ (see Fig.~\ref{fig:6}a), we determined the scale heights for
pulsars at $|z_\mathrm{p}| < 0.9\kpc$. There are 130 pulsars in this
sample.

\begin{figure}[htb]
\includegraphics[bb = 136 89 543 700,angle=270,width=8.5cm]{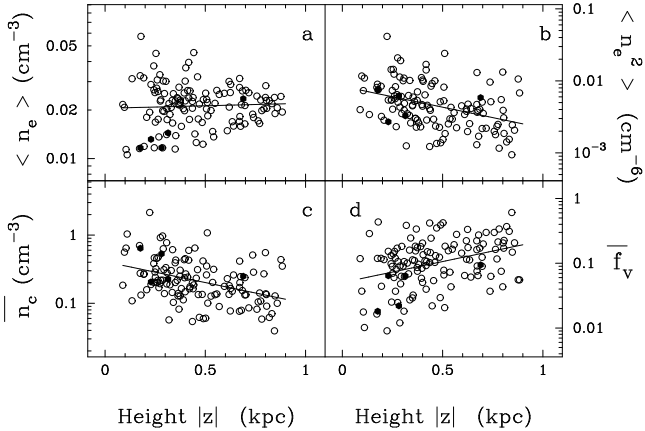}
\caption{Electron densities averaged along the line of sight
and average volume filling factor as function of height above
the Galactic plane, $|z|$.
{\bf (a)} average density $\avnel = \DM \sin |b|/|z|$.
{\bf (b)} average of the square of the density $\langle
\nel^2\rangle = \EM_\mathrm{p}\sin |b|/|z|$.
{\bf (c)} mean density in clouds $\ovnec = \EM_\mathrm{p}/\DM$.
{\bf (d)} average volume filling factor $\ovfilfac =
\DM^2 /(\EM_\mathrm{p} D)$. Full lines indicate the exponential fits
for $|z| < 0.9\kpc$ given in Table~\ref{tab:2}. Filled circles are
pulsars with parallactic distances. }
  \label{fig:7}
\end{figure}

Figure~\ref{fig:7} shows the distributions of the various densities and
$\ovfilfac$ in the $\ln Y-|z|$ plane. The exponential fits are
listed in Table~\ref{tab:2}.

\begin{table*}[htb]
\caption{Statistical relationships for $h=0.28\kpc$}
\label{tab:2}
\begin{tabular}{lllll@{$\pm$}lr@{$\pm$}ll@{$\pm$}lccc}
\hline\hline
\noalign{\medskip}
\multicolumn{1}{l}{$X$} &
\multicolumn{1}{l}{$Y$} &
\multicolumn{1}{l}{Function} &
\multicolumn{1}{l}{Kind} &
\multicolumn{2}{c}{} &
\multicolumn{2}{c}{} &
\multicolumn{2}{c}{corr.coeff.} &
\multicolumn{1}{c}{Student} &
\multicolumn{1}{c}{$|z|$} &
\multicolumn{1}{c}{N} \\
\multicolumn{1}{l}{} &
\multicolumn{1}{l}{} &
\multicolumn{1}{l}{} &
\multicolumn{1}{l}{of fit}&
\multicolumn{2}{c}{$a$}&
\multicolumn{2}{c}{$b$}&
\multicolumn{2}{c}{$r$}&
\multicolumn{1}{c}{$t$}&
\multicolumn{1}{c}{[kpc]} &
\multicolumn{1}{c}{} \\
\noalign{\medskip}
\hline
\noalign{\medskip}
$\DM$          &$\EM$  &$Y=aX^\mathrm{b}$   &bis.$^{1)}$
       &\pheins 0.052&0.020  &1.48&0.10  &0.65&0.06 &10.7 &all &157\\
$\DM$          &$\EM_\mathrm{p}$  & "       & "
       &\pheins 0.042&0.014  &1.47&0.09  &0.76&0.05 &14.5 & " & "\\
$\DM\cos|b|$   &$\EM_\mathrm{p}\cos|b|$  & "  & "
       &\pheins 0.055&0.015  &1.41&0.07  &0.85&0.04 &20.2 &"  &"\\
$\DM\sin|b|$   &$\EM_\mathrm{p}\sin|b|$  & "  & "
       &\pheins 0.168&0.027  &1.08&0.06  &0.44&0.07 &\pheins 6.1 &" &"\\
\noalign{\medskip}
$|z|$          &$\DM\sin|b|$   & "         &pow.$^{2)}$
       &22.1&1.3  &1.04&0.05  &0.89&0.04  &22.3 &$<0.9$ &130\\
$|z|$          &$\EM_\mathrm{p}\sin|b|$  & "  & "
       &\pheins 2.78&0.28  &0.46&0.11  &0.68&0.06  &10.6 &" &"\\
\noalign{\medskip}
$|z|$          &$\avnel$  &$Y=ae^{|z|/b}$   &exp.+$^{3)}$
       &\pheins 0.0205&0.0014 &\multicolumn{2}{c}{$14^{~...}_{~-8}$$^{4)}$}
       &0.66&0.07  &\pheins 9.8  &$<0.9$ &130\\
$|z|$          &$\langle\nel^2\rangle$   & "  &"\hspace{0.42cm}$-$
       &\pheins 0.0084&0.0012 &\multicolumn{2}{c}{$0.75^{+0.20}_{-0.13}$}
       &0.71&0.06  &11.4  &"  &"\\
$|z|$          &$\ovnec$    & "           &"\hspace{0.42cm}$-$
       &\pheins 0.407&0.059 &\multicolumn{2}{c}{$0.71^{+0.18}_{-0.12}$}
       &0.72&0.06  &11.8  &"  &"\\
$|z|$          &$\ovfilfac$  & "          &"\hspace{0.42cm}$+$
       &\multicolumn{2}{l}{$\pheins 0.0504^{+0.0095}_{-0.0080}$}
       &\multicolumn{2}{c}{$0.67^{+0.20}_{-0.13}$} &0.72&0.06
       &"  &"\\
\noalign{\medskip}
$\ovnec$       &$\ovfilfac$  &$Y=aX^\mathrm{b}$    &pow.
       &\pheins 0.0178&0.0015  &$-$1.11&0.04  &0.95&0.03 &33.4 &$<0.9$ &130\\
    & & & &\pheins 0.0184&0.0011  &$-$1.07&0.03  &0.94&0.03  &35.1 &all
         &157\\
$\ovnec$       &$L_\mathrm{e}$    & "       & "
       &35&6  &$-$0.82&0.10  &0.72&0.06  &11.8 &$<0.9$ &130\\
    & & & &35&5  &$-$0.86&0.06  &0.74&0.05  &13.6 &all    &157\\
\noalign{\medskip}
\hline
\noalign{\medskip}
\multicolumn{13}{l}{1) Bisector in $\log Y - \log X$ plane.
2) Regression line of $\log Y$ on $\log X$. 3) Regression line of $\ln
Y$ on $|z|$ with slope}\\
\multicolumn{13}{l}{$1/b$, where $b$ is the scale height $H$ in kpc;
+/$-$ means increases/decreases with $|z|$. All fits are ordinary
least-squares}\\
\multicolumn{13}{l}{fits with errors of one standard deviation. Fits with $t>3.1$ are
significant at the 3$\sigma$-level. 4) Error is undefined.}\\
\multicolumn{13}{l}{Units: $\DM$ in $\rm cm^{-3}\pc$, $\EM$ and
$\EM_\mathrm{p}$ in $\rm cm^{-6}\pc$, $\avnel$ and $\ovnec$ in $\rm
cm^{-3}$, $\langle\nel^2\rangle$ in $\rm cm^{-6}$, $|z|$
in kpc, $L_\mathrm{e}$ in pc.}\\
\end{tabular}
\end{table*}

As the model of Cordes \& Lazio (\cite{cordes+lazio02}) assumes that
$\avnel$ is constant, $\avnel$ is independent of $|z|$
(Fig.~\ref{fig:7}a). The spread in $\avnel$ is within a factor of 2 to
3 around the mean; the mean value at the midplane $\avnel_0$ of
$0.0205\pm 0.0014\cmcube$ is in good agreement with earlier
determinations for the solar neighbourhood  (Reynolds\
\cite{reynolds91b}; Weisberg et al.\ \cite{weisberg+80}). The spread
decreases steadily with increasing $|z|$. This is a real
$|z|$-effect and not just caused by a longer line of sight, because in
a plot of $\avnel - \mathrm{D} \cos |b|$ (not shown) the spread in
$\avnel$ remains the same up to a distance of 3~kpc, beyond which only
13 pulsars are located.

Figure~\ref{fig:7}b shows a decrease of $\langle\nel^2\rangle$ with
$|z|$ corresponding to a scale height $H$ of about $750\pc$.
The bottom panels of Fig.~\ref{fig:7} show the variations of $\ovnec$
(left) and $\ovfilfac$ (right) with $|z|$. The exponential fits show
that $\ovnec (|z|)$ decreases with $H\simeq 710\pc$, whereas $\ovfilfac
(|z|)$ increases with $H\simeq 670\pc$. Their midplane values are
$\overline{n}_\mathrm{c,0} = 0.41\pm 0.06\cmcube$ and
$\overline{f}_\mathrm{v,0} = 0.050\pm 0.009$, respectively (see
Table~\ref{tab:2}). Note that the near equality of the scale heights of
$\langle\nel^2\rangle$ and $\ovnec$, and of $\ovfilfac$ with opposite
sign, follows from the relations $\ovnec = \langle\nel^2\rangle /
\avnel$, $\ovfilfac = \avnel^2 / \langle\nel^2\rangle$ and the constant
value of $\avnel$ with $|z|$ (see Eqs. (1) to (4)).
The increase of $\ovfilfac$ with $|z|$ could cause the decreasing
spread in $\avnel$ with $|z|$ as long lines of sight through low-density
regions at high $|z|$ will give less variation in $\avnel$ than short
lines of sight through high-density regions at low $|z|$.

We conclude that up to $|z| = 0.9\kpc$ $\avnel$ the scale heights $H$
of $\langle\nel^2\rangle$, $\ovnec$ and $\ovfilfac$ are about 0.7~kpc
for this sample of pulsars at $60\degr <
\ell < 360\degr$ and $|b| > 5\degr$.

\subsection{Volume filling factors and $\ovnec$}
\begin{figure}[htb]
\includegraphics[bb = 136 89 537 624,angle=270,width=8.5cm]{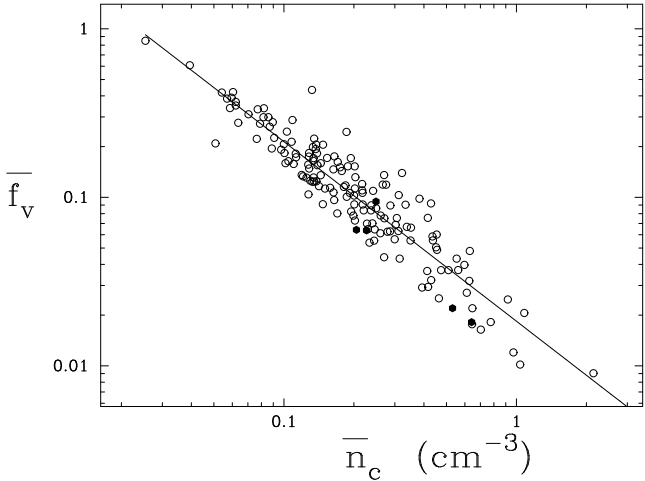}
\caption{Dependence of the average volume filling factor $\ovfilfac$ on
the mean density in clouds $\ovnec$ for the final sample of 157
pulsars. The full line represents the power-law fit $\ovfilfac (\ovnec )
= (0.0184\pm 0.0011) \ovnec^{~-1.07\pm 0.03}$. The correlation
coefficient $r = 0.94\pm 0.03$ and the Student-t = 35 (see
Table~\ref{tab:2}). Filled circles are pulsars with parallactic
distances.}
  \label{fig:8}
\end{figure}

In Fig.~\ref{fig:8} we present the dependence of the volume filling
factor $\ovfilfac$ on the mean electron density in clouds $\ovnec$ for
all pulsars in the sample (N = 157). The correlation is very good: the
power-law fit yields $\ovfilfac (\ovnec) = (0.0184\pm 0.0011)
\ovnec^{~-1.07\pm0.03}$ with a correlation coefficient $r = 0.94\pm0.03$
and $t=35$ (see Table~\ref{tab:2}). It nearly covers two orders of
magnitude in $\ovnec$ ($0.02-2\cmcube$) and $\ovfilfac$ ($0.9-0.009$).
For pulsars at $|z|<0.9\kpc$ the resulting fit is nearly the same.
While an inverse correlation between $\ovfilfac$ and $\ovnec$ is
expected from the near constancy of $\avnel = \ovfilfac \ovnec$,
the combination of $\DM$ and $\EM$ yields the values of $\ovnec$
and $\ovfilfac$ for which it holds.

The small spread of about a factor 2 in $\ovfilfac$ for a certain
$\ovnec$ is remarkable. This is due to two fortunate facts: 1. since
$\ovnec = \EM_\mathrm{p}/\DM$ and $\ovfilfac = \DM^2/\EM_\mathrm{p}D$,
$D$ and errors in $D$ cancel in both quantities (see Eqs. (1) to (4)),
2. any variation in $\EM_\mathrm{p}$, which causes the largest spread
in the various correlations, influences $\ovfilfac$ and $\ovnec$ by the
same factor but in opposite directions. Therefore points just slide up
or down along the distribution of points with slope $-1$. For the same
reason the corrections to EM have little effect on the $\ovfilfac
(\ovnec )$ relation. Without corrections we find $\ovfilfac (\ovnec ) =
(0.0193\pm0.0009) \ovnec^{~-1.04\pm0.03}$ for N~=~157, which  holds for
somewhat higher densities $(0.04 < \ovnec < 4\cmcube)$ than the
corrected relation.

The inverse dependence of $\ovfilfac$ on $\ovnec$ obtained here is
somewhat steeper than the dependence derived by Pynzar
(\cite{pynzar93}) and Berkhuijsen (\cite{elly98}). Their result,
$\ovfilfac \propto \ovnec^{~-0.7}$, was based on a large variety of
data, including classical \ion{H}{ii} regions, and the pulsar distances
were less well known. Without the classical \ion{H}{ii} regions Pynzar's
data indicate a slope closer to $-1$. We believe that our coherent
data set of 157 pulsars at $|b| > 5\degr$ more reliably represents
the inverse relationship between $\ovfilfac$ and $\ovnec$ for
the diffuse ionized gas. For mean cloud densities in the range $0.03 <
\ovnec < 2\cmcube$ we find volume filling factors of $0.8 > \filfac >
0.01$ (see Fig.~\ref{fig:8}).

\begin{figure}[htb]
\includegraphics[bb = 136 78 537 624,angle=270,width=8.5cm]{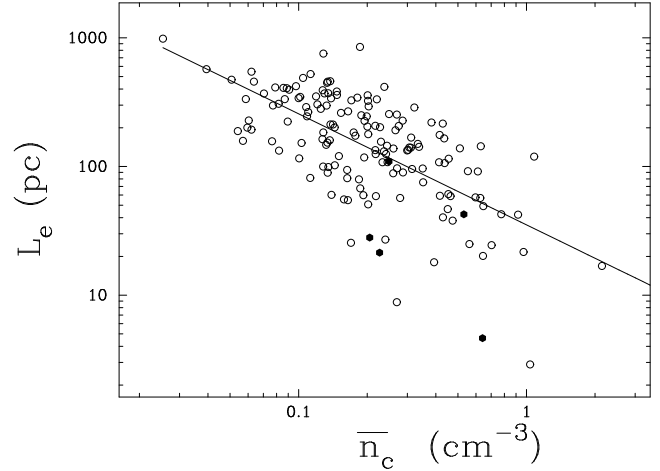}
\caption{Dependence of the total path length along the line of
sight through the ionized
regions, $L_\mathrm{e}$, on the mean density in these regions $\ovnec$.
The full line represents the power-law fit for the sample of 157 pulsars
(see Table~\ref{tab:2}). Filled circles are pulsars with parallactic
distances. }
  \label{fig:9}
\end{figure}

\subsection{Extent of ionized regions}
As the volume filling factor $\ovfilfac$ of the DIG increases with
decreasing mean density $\ovnec$, also an increase of the extent of the
ionized regions is expected. This relationship is shown in
Fig.~\ref{fig:9} where we plotted $L_\mathrm{e}$ as a function of
$\ovnec$ for the total sample of 157 pulsars. A power-law fit
yields $L_\mathrm{e} (\ovnec ) = (35\pm 5) \ovnec^{~-0.86\pm0.06}\pc$
with high significance. The extent of the ionized gas along the line of
sight towards the pulsars ranges from about 35~pc for $\ovnec
\simeq 1\cmcube$ to about 500~pc for $\ovnec \simeq 0.05\cmcube$.
Like in the case of $\ovfilfac (\ovnec )$, the slope of this
relationship is insensitive to errors in $\EM_\mathrm{p}$. The spread
in the distribution of $L_\mathrm{e}(\ovnec )$ is larger than in that
of $\filfac(\ovnec )$, because $L_\mathrm{e}$ still contains the
distance $D$.

Perhaps more interesting than the extent of the ionized regions towards
the pulsars, which depends on their latitude, is their extent in the
$|z|$-direction. This is best obtained from the relation $L_\mathrm{e}
\sin |b|(z) = \ovfilfac (z)|z|$. $L_\mathrm{e} \sin |b|$ grows from
about 6~pc for $|z| = 100\pc$ to 53~pc for $|z| = 500\pc$ and about
200~pc near $|z| = 1\kpc$. The corresponding mean densities in the
clouds, calculated from $\ovnec (z)$ given in Table~\ref{tab:2}, are
$0.35\cmcube,\ 0.20\cmcube$ and $0.12\cmcube$, respectively.

\subsection{Dependence of various relationships on $h$}
The results presented in Table~\ref{tab:2} were obtained using
$h=0.28\kpc$ to correct the observed emission measures for contributions
from beyond the pulsar distance (see Eq.~6). However, in Sect.~4.1 we
find that $h$ could be as high as $0.5\kpc$. To see how $h>0.28\kpc$
influences the values in Table~\ref{tab:2}, we repeated the fitting
procedures for $h=0.36\kpc$ and $h=0.5\kpc$.

The correlations $Y(X)$ corresponding to $\EM_\mathrm{p} (\DM )$,
$\EM_\mathrm{p} \cos |b| (\DM\cos |b|)$ and $\ovfilfac (\ovnec )$
remained the same to within 1 s.d. from the values in
Table~\ref{tab:2}, while those of $\EM_\mathrm{p} \sin |b| (\DM \sin
|b|)$ and $\EM_\mathrm{p} \sin |b| (|z|)$ stayed within 2 s.d. from the
values in Table~\ref{tab:2}. Only the scale heights $H$ and the
midplane values $Y_0(h)$ of $\langle \nel^2 \rangle$, $\ovnec$ and
$\ovfilfac$ changed by $\ga 2$ s.d. if $h= 0.5\kpc$ is used. For $0.25
\le h \le 0.5\kpc$ these values can be found from

\begin{equation}
H(h) = 1.8 h + 0.24\kpc \quad \mbox{for}\ \langle \nel^2 \rangle \ .
\end{equation}
The scale heights of $\ovnec$ and $\ovfilfac$ are $0.95 H(h)$ and $0.89
H(h)$, respectively.

\begin{eqnarray}
Y_0(h) & = & 1.04 Y_0 (0.28) * 0.28/h \quad \mbox{for}\ \langle
\nel^2\rangle\ \mbox{and}\ \ovnec\ , \nonumber \\
Y_0(h) & = & 1.04 Y_0 (0.28) * h/0.28 \quad \mbox{for}\ \ovfilfac \ ,
\end{eqnarray}
where $Y_0 (0.28) = a$ in Table~\ref{tab:2}.

The general relationship between $H$ and $h$ is given by

\begin{eqnarray}
\langle \nel^2 \rangle (z) & = & \frac{\EM_\mathrm{p} \sin |b|(z)}{|z|}
  = \frac{1}{|z|} \int^{|z|}_0 \nel^2 (0) e^{-|z|/h} dz \nonumber \\
 & \simeq & \langle \nel^2 \rangle_0 e^{-|z|/H} \ .
\end{eqnarray}
As $\langle \nel^2 \rangle (z)$ is not a purely exponential function,
the last expression is only an approximation. Similar relations hold for
$\ovnec$ and $\ovfilfac$. We checked how well this approximation
describes the observations by simulating a data set of $\langle \nel^2
\rangle (z)$ calculated from the integral between $|z|=0$ and $0.9\kpc$
divided by $|z|$ for $h=0.28\kpc$, $\nel^2 (0) = 1\cmsix$, and fitting
an exponential to this (noiseless) distribution, exactly as we did in
Sect.~3.2. We found $H=0.75\pm 0.02\kpc$ and $\langle \nel^2 \rangle_0 =
0.94\pm 0.01\cmsix$. The errors and the decrease of 6\% in $\langle
\nel^2 \rangle_0$ are caused by the curvature of $\ln
\langle\nel^2\rangle (z)$ versus $|z|$. Thus the simulation yields the
same scale height $H$ as derived from the (noisy) observations (see
Table~\ref{tab:2}) and indicates that the midplane value in
Table~\ref{tab:2} may be 6\% low, which is within the observational
errors.

\section{Discussion}

In this section we focus on the vertical structure of the electron
layer. We derive the scale height $h$ of the local electron density in
Sect.~4.1, compare this with other scale heights in Sect.~4.2 and
discuss the relationship between $\ovfilfac$ and $\ovnec$ in Sect.~4.3.

\subsection{Vertical structure of the DIG}

\begin{figure}[htb]
\includegraphics[bb = 136 97 552 624,angle=270,width=8.5cm]{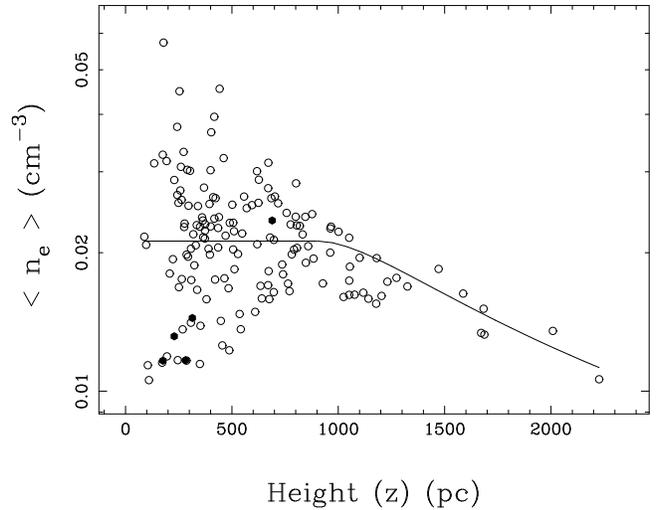}
\caption{Dependence of the average electron density along the line
of sight $\avnel = \DM \sin |b|/|z|$ on height above the Galactic plane
$|z|$ for the sample of 157 pulsars.
The full horizontal line for $|z| < 0.9\kpc$ is at $\avnel =
0.0213\cmcube$. The continuation to larger $|z|$ shows the
values expected (Eq.~(11) divided by $|z|$) for a true scale height
$h=280\pc$ and maximum local filling factor $f(z_1) = 1$ at $|z_1| =
0.9\kpc$. See Sect.~4.1 for explanations. }
  \label{fig:10}
\end{figure}

The variation of $\avnel$ with $|z|$ for all pulsars in our sample
(N=157) is shown in Fig.~\ref{fig:10}. Below $|z| \simeq 0.9\kpc$
$\avnel$ is nearly constant (see also Fig.~\ref{fig:7}a) but beyond this
height $\avnel$ starts decreasing. This variation reflects the linear
increase of $\DM \sin |b|$ up to $|z| \simeq 0.9\kpc$ and its levelling
off at higher $|z|$ observed for various samples of pulsars at known
distances (see also Fig.~\ref{fig:6}a). Such a behaviour of $\DM \sin
|b|(z)$ is expected if the local electron density $\nel (z)$ decreases
exponentially with a scale height of about 1~kpc (Reynolds\
\cite{reynolds91b}; Bhattacharya \& Verbunt\ \cite{bhatta+verbunt91};
Nordgren et al.\  \cite{nordgren+92}; G\'omez et al.\ \cite{gomez+01};
Cordes \& Lazio\  \cite{cordes+lazio03}) which was also adopted by
Cordes \& Lazio (\cite{cordes+lazio02}) for their model. However, in
an analysis of the dependence of $\DM \sin |b|$ on $|z|$ of about 70
pulsars with independent distance estimates, Cordes \& Lazio
(\cite{cordes+lazio03}) noted that this dependence may be due to a
constant local electron density up to $|z| \simeq 1\kpc$ and that
the electron layer may be bounded near this height. Here we give an
alternative explanation of the dependence of $\DM \sin |b|$ on $|z|$.

The key observation is the constancy of $\avnel$ for $|z| \la
0.9\kpc$. This implies that the local electron density $\nel (z)$ is
also constant and that the local filling factor $f(z)$ and local
density in clouds $\nec(z)$ are inversely correlated.
We expect $\nec (z)$ to decrease with increasing $|z|$. This is
indicated by the decrease of emission measure observed for the Perseus
arm (Haffner et al.\ \cite{haffner+99}) and external, edge-on galaxies.
Hence, $f(z)$ increases with $|z|$. Beyond $|z| \simeq 0.9\kpc$ $\DM
\sin |b|(z)$ flattens and $\avnel(z)$ starts decreasing (see
Fig.~\ref{fig:10}). We suggest that this occurs because near $|z| =
0.9\kpc$ $f(z)$ reaches a maximum whereas $\nec(z)$ continues to
decrease beyond this height. $\DM \sin |b|(z)$ may then be
described in the following way:
\begin{eqnarray}
\DM \sin |b|(z) & = & \int_0^{|z|} \nel (z) dz =
                      \int_0^\mathrm{|z|} f(z) \nec(z) dz \nonumber\\
& = & \int_0^\mathrm{|z|} f_0\, e^\mathrm{|z|/h_f}\ n_\mathrm{c,0}\
       e^\mathrm{-|z|/h_n} dz \  ,
\end{eqnarray}
where we assumed that $f(z)$ and $\nec(z)$ are exponentials with
midplane values $f_0$ and $n_\mathrm{c,0}$, and scale heights
$h_\mathrm{f}$ and $h_\mathrm{n}$, respectively.

We know that for $|z| < |z_1| \simeq 0.9\kpc$ $\DM \sin |b|(z) = \avnel
|z|$. Hence $h_\mathrm{f} = h_\mathrm{n} = h$, $f_0 n_\mathrm{c,0} =
\nel (0) = \avnel$ and $f(z) < f(z_1)$ where $f(z_1)$ is the maximum
value. For $|z| \ge |z_1|$ we have
\begin{equation}
\DM \sin |b|(z) = \avnel |z_1| + f(z_1) \int^{|z|}_{|z_1|}
   n_\mathrm{c,0}\, e^{-|z|/h} dz \ .
\end{equation}
Thus for very large $|z|$ $\DM \sin |b|(z)$ reaches a maximum
\begin{equation}
\DM \sin |b|_\mathrm{max} = \avnel |z_1| + f(z_1) n_\mathrm{c,0} h\,
e^{-|z_1|/h} \ ,
\end{equation}
where $|z_1|$ is the height at which the turn-over in $\DM \sin |b|(z)$
starts and $f(z_1) = f_0\, e^{|z_1|/h}$.

Similarly, we may describe the emission measure perpendicular to the
Galactic plane as
\begin{displaymath}
\EM_\mathrm{p} \sin |b|(z) = \int^{|z_\mathrm{p}|}_0 \nel^2 (z) dz
 = \int_0^{|z_\mathrm{p}|} f(z) \nec^2 (z) dz \ .
\end{displaymath}
Since $f(z) \nec(z) = f_0 n_\mathrm{c,0}$ for $|z| < |z_1|$ and $f(z) =
f(z_1)$ for $|z| \ge |z_1|$, we have
\begin{eqnarray*}
\EM_\mathrm{p} \sin |b|(z) & = & f_0 n_\mathrm{c,0}
\int_0^{|z<z_1|} n_\mathrm{c,0}\, e^{-|z|/h} dz \nonumber\\
& + & f(z_1) \int_{|z_1|}^{|z>z_1|} n^2_\mathrm{c,0}\,
 e^{-2|z|/h} dz \ .
\end{eqnarray*}
Thus the maximum emission measure through the electron layer is
\begin{eqnarray}
\EM_\mathrm{p} \sin |b|_\mathrm{max} & = &\avnel n_\mathrm{c,0} h
 \left( 1 - e^{-|z_1|/h}\right) \nonumber\\
& + & f(z_1) \frac{n^2_\mathrm{c,0}h}{2}\,
 e^{-2|z_1|/h} \ .
\end{eqnarray}

\begin{table*}[htb]
\caption{Possible range of true scale height $h$}
\label{tab:3}
\begin{tabular}{lcccccc}
\hline\hline
\noalign{\medskip}
\multicolumn{1}{l}{$f(z_1)$}&
\multicolumn{1}{c}{$h$}&
\multicolumn{1}{c}{$n_\mathrm{c0}$}&
\multicolumn{1}{c}{$f_0$}&
\multicolumn{1}{c}{$|z_1|$}&
\multicolumn{1}{c}{$\DM\sin|b|(z_1)$}&
\multicolumn{1}{c}{$\DM\sin|b|_\mathrm{max}$}\\
\multicolumn{1}{l}{}&
\multicolumn{1}{c}{[pc]}&
\multicolumn{1}{c}{[$\rm cm^{-3}$]}&
\multicolumn{1}{c}{}&
\multicolumn{1}{c}{[pc]}&
\multicolumn{1}{c}{[$\rm cm^{-3}\pc$]}&
\multicolumn{1}{c}{[$\rm cm^{-3}\pc$]}\\
\noalign{\medskip}
\hline
\noalign{\medskip}
1$^{1)}$   &250       &0.600  &0.0355  &836\phnote  &17.7  &23.0\\
        &{\bf 280} &0.536  &0.0397  &904\phnote  &19.2  &25.1\\
        &310       &0.484  &0.0440  &970\phnote  &20.6  &27.1\\
\noalign{\smallskip}
0.5$^{1)}$ &320       &0.469  &0.0454  &767\phnote  &16.3  &23.2\\
        &{\bf 360} &0.417  &0.0511  &821\phnote  &17.5  &25.2\\
        &400       &0.375  &0.0568  &870\phnote  &18.5  &27.1\\
\noalign{\smallskip}
0.46    &330       &0.455  &0.0469  &750$^{1)}$  &16.0  &23.1\\
0.36    &{\bf 420} &0.357  &0.0596  &"           &"     &25.0\\
0.31    &520       &0.288  &0.0738  &"           &"     &27.1\\
\noalign{\medskip}
\hline
\noalign{\medskip}
\multicolumn{7}{l}{$^{1)}$ assumed value. See Sect.~4.1 for
explanations}\\
\end{tabular}
\end{table*}

\noindent
This is equal to the absorption-corrected emission measure
$\EM_\mathrm{c} \sin |b|$ discussed in Sect.~2.3. For $|z| > |z_1|$ we
expect little increase of the emission measure because the second term
in Eq.~(13) is much smaller than the second term in Eq.~(12) and even
the dispersion measures increase by $\la25$\% beyond $|z| = 0.9\kpc$
(see Fig.~\ref{fig:6}a). Therefore we assume that at $|z| = |z_1|$ the
maximum value is nearly reached and reduce Eq.~(13) to
\begin{equation}
\EM_\mathrm{p} \sin |b|_\mathrm{max} = \avnel n_\mathrm{c,0} h \ .
\end{equation}

We can now evaluate the scale height $h$ by combining Eqs. (12) and (14)
with known data. For $\avnel$ we take the mean value of the slightly
sloping function $\avnel(z)$, $\avnel = f_0 n_\mathrm{c,0} = 0.0213\pm
0.0002\cmcube$. The average value of the absorption-corrected emission
measures $\overline{\EM_\mathrm{c}\sin|b|} = 3.2\pm 0.2\cmsix\pc$ (see
Sect.~2.3), which gives $n_\mathrm{c,0}h = 150\pm 10\cmcube\pc$. From
Fig.~\ref{fig:6}a we obtain $\DM\sin|b|_\mathrm{max} = 25\pm
2\cmcube\pc$ and $z_1 \simeq 0.9\kpc$. We first take $f(z_1) = f_0\,
e^{|z_1|/h} = 1$ which gives a lower limit to $h$. For an assumed value
of $h$ we then calculate $n_\mathrm{c,0}$, $f_0$, $|z_1|$, $\DM
\sin|b|(z_1)$ and $\DM\sin|b|_\mathrm{max}$. We found $|z_1|$ and
$\DM\sin|b|_\mathrm{max}$ to be in the correct range for $250 < h <
310\pc$. Repeating the procedure for a more realistic value of $f(z_1) =
0.5$ yielded a somewhat larger scale height: $320 < h < 400\pc$
(see Table~\ref{tab:3}). The height $|z_1|$ decreases when $h$
increases. The lowest value of $|z_1|$ consistent with the data in
Fig.~\ref{fig:6}a is $\simeq 750\pc$. Assuming this value gives an
upper limit to $h$ in the range $330 < h < 520\pc$ and a maximum
filling factor of $0.46 > f(z_1)>0.31$. We conclude that the scale
height of the local electron density in the DIG is in the range $250 <
h \la 500\pc$ and that the maximum filling factor $f(z_1) > 0.3$.
Because the lower limit to $h$ is best constrained, we used $h =
280\pc$ to correct the emission measures for contributions from beyond
the pulsars in Sect.~2.3. How a larger value of $h$ changes the results
in Table~\ref{tab:2} is described in Sect.~3.5.

In Fig.~\ref{fig:10} the expected variation of $\avnel(z)$ is shown for
$h = 280\pc$ and $f(z_1) = 1$. The agreement with the data at $|z| >
0.9\kpc$ is very good. With $f(z_1) = 0.5$ $\avnel(z)$ is constant up
to $|z_1| = 820\pc$ and then decreases more slowly to the same value at
$|z| = 2\kpc$ as for $f(z_1) = 1$. Also for $|z_1| = 750\pc$ and
$h=500\pc$ the line is nearly identical to those of the two other cases.

For an electron layer with constant $\avnel$ up to 1~kpc and then a
sudden cutoff, as proposed by Cordes \& Lazio (\cite{cordes+lazio03}),
$\avnel(z) = \avnel (|z| = 1\kpc) / |z|$ for $|z| > 1\kpc$. This
leads to a somewhat faster decrease in $\avnel$ at high $|z|$ than is
observed. A more gradual end of the layer seems more realistic and
might be in better agreement with the data.

Above we have described that $\avnel(z)$ will be constant as long as the
local filling factor and electron density in clouds are inversely
correlated. This correlation breaks down when the filling factor has
reached a maximum value of $\ga35$\% near $|z| = 0.8\kpc$, whereas the
electron density in clouds continues to decrease to higher $|z|$. This
picture is in good agreement with the data in Fig.~\ref{fig:10} and
shows that the electron layer does not need to be cut off. Why the
filling factor reaches a maximum at high $|z|$ is an interesting
question. Pressure balances must play an important role and model
calculations will be needed to better understand what is going on near
$|z| = 1\kpc$ (see also Sect.~4.3.1).

\subsection{Scale heights and column densities}

In the foregoing section we showed that at heights above the plane $\la
1\kpc$ the scale height of the emission measure equals the scale height
$h$ of the local electron density in clouds $\nec$. This happens because
the local volume filling factor $f$ increases with the same scale height
leading to $\nel (z) = \avnel \simeq \mbox{constant}$ (see Eq. 13).
However, at larger heights the filling factor remains constant at the
maximum value $f(z_1)$ and the scale height of the emission measure
becomes $h/2$, the scale height of $\nec^2$. Unfortunately the emission
measure scale height is difficult to observe and the available
estimates usually refer to $|z| \la 1\kpc$.

A lower limit to $h$ comes from radio recombination lines observed
between $\ell = 332\degr$ and $83\degr$ near $b=0\degr$ (Roshi \&
Anantharamaiah\ \cite{roshi+anan00}). These lines originate from
low-density ($1-10\cmcube$) ionized gas in envelopes of normal
\ion{H}{ii} regions. The scale height of this gas is $\ga 100\pc$.
H$\alpha$ observations of M\,31 (Walterbos \& Braun\
\cite{walterbos+braun94}) and other nearby galaxies (Beckman et al.\
\cite{beckman+02}) show that the DIG is concentrated around the
\ion{H}{ii} regions in the spiral arms. The latter authors argue that
electrons leaking from bounded \ion{H}{ii} regions can travel very large
distances and could be the origin of the DIG at high $|z|$.

Reynolds (\cite{reynolds86}) estimated the scale height from a
comparison of  H$\alpha$ intensities towards the poles and a mean value
in the Galactic plane. After correction for  absorption and scattering
he obtained a scale height of about 250~pc (corrected to $R_{\sun} =
8.5\kpc$), within the range of our value of $250 < h < 500\pc$,
although he used a very different method.

Berkhuijsen et al. (\cite{elly+97}) estimated the scale height of the
thermal radio continuum emission at 1.4~GHz, which is unaffected by
absorption, using the data of Reich \& Reich (\cite{reich+reich88}). The
latitude extent of the thermal emission at galactocentric distance
$R\simeq 4\kpc$ indicates a scale height of about 350~pc (after
correction for the size of the telescope beam).

From H$\alpha$ observations in the direction $125\degr < \ell <
152\degr$ Haffner et al. (\cite{haffner+99}) derived a scale height of
the emission measure of $500\pm50\pc$ for distances between 700~pc and
1750~pc from the Galactic plane in the Perseus arm. Because of various
line-of-sight effects the observed emission measures are dominated by
gas at heights smaller than the geometrical $z$-distances and the
authors note that the derived scale height will be too large. This
scale height of $\la 500\pc$ will represent a mixture of the true scale
height $h$ expected for $z < z_1 \simeq 1\kpc$ and $h/2$ expected for
$z>z_1$, hence $\la 500\pc < h \la 1000\pc$. We note that beyond the
Galactic solar radius the scale height of the gas increases. Our value
of $250 < h \la 500\pc$ for $|z|<|z_1|$ near the Sun would increase to
$300 < h \la 600\pc$ at the radius of $\sim 10\kpc$ of the Perseus arm
at $\ell \simeq 135\degr$ (Merrifield\ \cite{merrifield92}). Thus if
the scale height derived for the Perseus arm is more representative for
$z<z_1$ than for $z>z_1$ it is consistent with our value obtained for
the solar neighbourhood, otherwise it would be larger than expected
from the value near the Sun.

In view of all these data, the value of the true scale height derived
by us, $250 < h \la 500\pc$ for $|z|< |z_1| \simeq 1\kpc$ seems quite
plausible for diffuse ionized gas with mean densities in clouds of
$0.03 < \ovnec < 2\cmcube$ (see Figs.~\ref{fig:7}c and \ref{fig:8}).

\begin{figure}[htb]
\includegraphics[bb = 112 128 492 637,width=7cm,clip=]{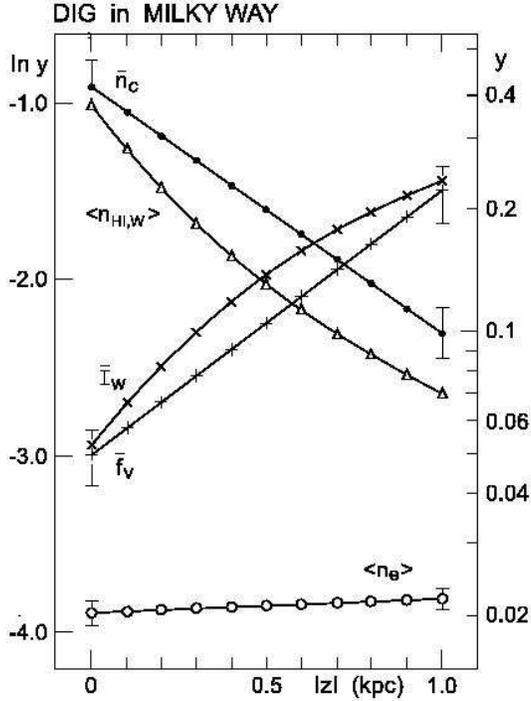}
\caption{Various quantities averaged along $|z|$ as a function of height
above the Galactic plane. Straight lines are exponential fits given in
Table~\ref{tab:2}. Dots -- mean electron density in ionized clouds
$\ovnec = \EM_\mathrm{p}/\DM$. Circles -- average electron density
$\avnel = \DM \sin |b|/|z|$. Triangles -- average density of warm
diffuse \ion{H}{i}, $\langle n_\mathrm{HI,w}\rangle = N(HI,w)/|z|$
taken from Diplas \& Savage (\cite{diplas+savage94}). Plusses --
average volume filling factor $\ovfilfac = \DM^2/(\EM_\mathrm{p} D)$.
Crosses -- average degree of ionization $\oviw = \avnel /(\avnel +
\langle n_\mathrm{HI,w}\rangle )$.
$\EM_\mathrm{p}$ was obtained with $h=280\pc$. With $h=500\pc$ the
curves of $\ovnec$ and $\ovfilfac$ are flatter than shown.
See Sect.~4.2 for explanations.
}
  \label{fig:11}
\end{figure}

In Fig.~\ref{fig:11} we plotted the variations of $\avnel$,
$\ovnec$ and $\ovfilfac$ with distance from the plane $|z|$ between 0
and 1~kpc calculated from the fitted exponentials in Table~\ref{tab:2}
which are based on $h=280\pc$. For $h=500\pc$ the curves of
$\ovnec$ and $\ovfilfac$ are flatter, because $\ovnec (0)$ drops from
$0.41\cmcube$ to $0.24\cmcube$ and $\ovfilfac (0)$ increases from 0.050
to 0.084, while the values at $|z| = 1\kpc$ do not change significantly.
We also plotted the average density along the line of sight of the
warm, diffuse \ion{H}{i} gas, $\langle n_\mathrm{HI,w}\rangle =
N(\mathrm{HI,w})/|z|$, derived from the work of Diplas \& Savage
(\cite{diplas+savage94}). This decreases faster than $\ovnec$ because
the true exponential scale height $h$ is only $195\pm 5\pc$ which
corresponds to $H=560\pm 20\pc$. Combining this with the data for
$\avnel$ yields average degrees of ionization of the diffuse gas,
$\oviw = \avnel / (\avnel + \langle n_\mathrm{HI,w}\rangle)$, between
$0.053\pm 0.004$ at the midplane to $0.24\pm 0.02$ at $|z| = 1\kpc$. It
is remarkable that $\oviw (z)$ and $\ovfilfac (z)$ are so similar.

Although in Fig.~\ref{fig:11} $\langle n_\mathrm{HI,w}\rangle$ is
everywhere larger than $\avnel$, the local density of warm \ion{H}{i},
$n_\mathrm{HI,w}$, drops below the (constant) local electron density
$\nel = f_0 n_\mathrm{c,0} = \avnel$ at $|z| \simeq 560\pc$ (see
Table~\ref{tab:4}). This height is close to the result of Reynolds
(\cite{reynolds91b}), who found that the total electron density becomes
larger than the total \ion{H}{i} density at $|z| \simeq 700\pc$. This
implies that above this height the local degree of ionization
$I_\mathrm{w}$ is larger than 0.5.

The column densities towards $|z| = 1\kpc$ of $\avnel$ and $\langle
n_\mathrm{HI,w}\rangle$ are $N(\avnel) = \DM \sin|b| = (0.68\pm 0.04)
10^{20}\, \mathrm{cm}^{-2}$ and $N(\mathrm{HI,w}) = (2.19\pm 0.06)
10^{20}\, \mathrm{cm}^{-2}$. They are similar to the lower values given
by Reynolds (\cite{reynolds91b}). This is understandable because our
sample of pulsars traces the diffuse gas at $|b| > 5\degr$ located in
mainly interarm regions.

\begin{table}[htb]
\caption{Local degree of ionization and volume filling factor}
\label{tab:4}
\begin{tabular}{rlllllll}
\hline\hline
\noalign{\medskip}
\multicolumn{1}{c}{$|z|$}&
\multicolumn{1}{l}{$n_\mathrm{HI,w}$}&
\multicolumn{2}{c}{$\nel / n_\mathrm{HI,w}$$^{1)}$}&
\multicolumn{2}{c}{$I_\mathrm{w}$$^{1)}$}&
\multicolumn{2}{c}{$f^{1)}$}\\
\multicolumn{1}{c}{[pc]}&
\multicolumn{1}{l}{[$\rm cm^{-3}$]}&
\multicolumn{2}{c}{}&
\multicolumn{2}{c}{}&
\multicolumn{1}{l}{}&
\multicolumn{1}{l}{}\\
\noalign{\medskip}
\hline
\noalign{\medskip}
0      &0.366  &\multicolumn{2}{c}{0.058}  &\multicolumn{2}{c}{0.055}
   &0.04  &0.07\\
200    &0.131  &\multicolumn{2}{c}{0.16\pheins}
   &\multicolumn{2}{c}{0.14\pheins} &0.08  &0.11\\
400    &0.047  &\multicolumn{2}{c}{0.45\pheins}
   &\multicolumn{2}{c}{0.31\pheins}   &0.17  &0.16\\
600    &0.017  &\multicolumn{2}{c}{1.25\pheins}
   &\multicolumn{2}{c}{0.56\pheins}   &0.34  &0.24\\
750    &0.0078 &\multicolumn{2}{c}{2.7\phelf}
   &\multicolumn{2}{c}{0.73\pheins}   &0.58  &0.32\\
800    &0.0061 &3.5  &3.2   &0.78  &0.76   &0.69  &0.32\\
900    &0.0036 &5.9  &4.4   &0.86  &0.81   &0.99  &0.32\\
1000   &0.0022 &6.8  &5.9   &0.87  &0.85   &1     &0.32\\
\noalign{\medskip}
\hline
\noalign{\medskip}
\multicolumn{8}{l}{$^{1)}$ Left-hand values for
   $h=280\pc$, right-hand values for}\\
\multicolumn{8}{l}{$h=500\pc$}\\
\multicolumn{8}{l}{$n_\mathrm{HI,w}(z) = 0.366 \exp (-|z|/195\pc)$
   (Diplas \& Savage\ \cite{diplas+savage94});}\\
\multicolumn{8}{l}{$I_\mathrm{w} = \nel /(\nel + n_\mathrm{HI,w})$;}\\
\multicolumn{8}{l}{$\nel (|z|<|z_1|) = f_0 n_\mathrm{c0} =
   0.021\cmcube$ and $\nel (|z|\ge|z_1|)$ =}\\
\multicolumn{8}{l}{$f(z_1) n_\mathrm{c}(z)$, where $f(z) = f_0 \exp
   (|z|/h)$ and}\\
\multicolumn{8}{l}{$\nec (z) = n_\mathrm{c0} \exp (-|z|/h)$.}\\
\multicolumn{8}{l}{$h=280\pc$: $|z_1|=900\pc$, $f(z_1) = 1$,
    $f_0=0.04$,}\\
\multicolumn{8}{l}{$n_\mathrm{c0} = 0.54\cmcube$}\\
\multicolumn{8}{l}{$h=500\pc$: $|z_1|=750\pc$, $f(z_1) = 0.32$,
    $f_0=0.07$,}\\
\multicolumn{8}{l}{$n_\mathrm{c0} = 0.30\cmcube$}\\
\end{tabular}
\end{table}

\subsection{Volume filling factors}
In this section we discuss the dependencies of $\ovfilfac$ on
$|z|$ and $\ovnec$ derived in Sect.~3.2 and possible implications.

\subsubsection{Dependence of $\ovfilfac$ on $|z|$}
Figures~\ref{fig:7}d and \ref{fig:11} show that the volume filling
factor averaged along the line of sight increases considerably with
height above the Galactic plane. Looking through the electron layer to
$|z| = 1\kpc$, $\ovfilfac = 0.22\pm 0.04$ for $h=280\pc$ and $\ovfilfac
= 0.25\pm 0.04$ for $h=500\pc$, values
amazingly close to the estimate of $\ovfilfac \ga 0.2$ for the full
layer made by Reynolds (\cite{reynolds91a}) based on dispersion
measures and emission measures towards 4 pulsars in globular clusters.
But this agreement is accidental. Using the method of Kulkarni \& Heiles
(\cite{kulkarni+heiles88}), Reynolds combined the mean values of
$\DM \sin |b|$ and $\EM \sin |b|$ observed for the 4 pulsars with values
of $\avnel_0$ and $\langle\nel^2\rangle_0$ obtained from other sources
and did not correct $\EM$ for absorption. The combination of data from
different sources may introduce errors. Furthermore, as Reynolds noted,
the large intrinsic spread in $\DM$ and $\EM$ and the small number of
pulsars make this estimate rather uncertain.

The average ionization degree $\oviw$ increases in nearly the same way
as $\ovfilfac$ (see Fig.~\ref{fig:11}). If thermal pressure balance
exists in the DIG both quantities would increase if the temperature
rises with height above the plane. Indications of a temperature
increase with $|z|$ have been reported by Haffner et al.
(\cite{haffner+99}) for the Perseus arm at $|z| > 700\pc$. Reynolds et
al. (\cite{reynolds+98}) found that the ionization ratio
$\nel/n_\mathrm{HI}$ is sensitive to temperature. We show this
ratio as a function of $|z|$ in Table~\ref{tab:4}. Since the filling
factor of \ion{H}{i} is not known, we calculated the ratio of the local
mean densities $\nel(z) = f(z) \nec (z)$ (see Sect.~4.1)
and $n_\mathrm{HI,w}(z) = 0.366 \exp (-|z|/195\pc)$ (Diplas \& Savage\
\cite{diplas+savage94}). Between $|z| = 0$ and $|z| = 1\kpc$ the
ionization ratio increases by more than a factor 100, reaching 1 at
$|z| = 560\pc$. Thus it seems possible that the temperature in the DIG
indeed increases with growing distance to the Galactic plane.

In Table~\ref{tab:4} we also list the local degree of ionization
$I_\mathrm{w} = \nel /(\nel + n_\mathrm{HI,w})$ and the local filling
factors $f(z)$ for $h=280\pc$ and $h=500\pc$ derived in Sect.~4.1.
They all increase up to $|z|=750\pc$, where $f(z)$ reaches the
maximum value of 0.32 if $h=500\pc$. At this height $I_\mathrm{w} =
0.73$. This high value suggests that the maximum value of $f(z)$
discussed in Sect.~4.1 occurs when the gas becomes nearly fully
ionized. The height above the Galactic plane where this happens is then
indicated by the $|z|$-distance at which $\DM\sin|b|(z)$ starts turning
over (see Fig.~\ref{fig:6}a).

\subsubsection{Dependence of $\ovfilfac$ on $\ovnec$}
The inverse correlation between $\ovfilfac$ and $\ovnec$, shown in
Fig.~\ref{fig:8}, is the tightest relationship of the 12 cases listed
in Table~\ref{tab:2}. As explained in Sect.~3.3, $\ovfilfac (\ovnec )$
is independent of distance and insensitive to errors in emission
measure.

Because both $\ovnec$ and $\ovfilfac$ vary with distance from the
Galactic plane (see Figs.~\ref{fig:7}c and \ref{fig:7}d), their
correlation could just be a $|z|$-effect. However, an inverse
relationship is also expected from the constancy of $\avnel$ with
$D\cos|b|$, the distance component parallel to the plane: $\avnel(0) =
0.0206\pm 0.0009\cmcube$ and the slope of the exponential fit in the
$\ln \avnel - D\cos|b|$ plane of $-0.005\pm 0.024$, i.e. effectively
zero.

Roshi \& Anantharamaiah (\cite{roshi+anan01}) made a survey of
radio recombination lines of the inner Galaxy at low latitudes. They
derived filling factors of $\la 0.01$ for extended regions of diffuse
ionized gas with densities of 1--$10\cmcube$. Their values are in good
agreement with the extension of our relationship in Fig.~\ref{fig:8}.

We conclude that the inverse relationship between $\ovfilfac$
and $\ovnec$ seems to hold everywhere in the DIG at least for the
density range $0.01 < \ovnec < 10\cmcube$. This general validity shows
that it describes a basic property of the DIG, possibly even of the
entire ISM (Berkhuijsen\ \cite{elly99}). Cordes et al.
(\cite{cordes+85}) estimated a filling factor of $10^{-4.0\pm 0.3}$ for
clumps of about 1~pc size causing scattering of pulsar signals. The
relationships $\ovfilfac (\ovnec )$ and $L_\mathrm{e}(\ovnec )$ in
Table~\ref{tab:2} give $\ovfilfac = 1.3\, 10^{-4}$ and $L_\mathrm{e} =
0.8\pc$ for $\ovnec = 100\cmcube$, so these clump properties also seem
to follow $\ovfilfac \propto \ovnec^{~-1}$ derived for the DIG.
Furthermore, Gaustad \& Van Buren (\cite{gaustad+vanburen93}) found
a similar relationship between filling factor and mean density for
clouds of diffuse dust within $400\pc$ from the Sun: $\overline{f}
(\overline{n}_\mathrm{d}) = (0.056\pm 0.020)
\overline{n}_\mathrm{d}^{\hspace{0.2em} -1.25\pm0.40}$ for $1 <
\overline{n}_\mathrm{d} < 10\cmcube$. For dust densities
$\overline{n}_\mathrm{d} = 1\cmcube$ and $10\cmcube$ this yields dust
filling factors of 0.06 and 0.003, respectively,  close to the filling
factors of the DIG for the same densities $\ovnec$.

Because errors in emission measure hardly influence the
$\ovfilfac (\ovnec )$-relation, we also made some fits without any
corrections to EM. For the sample $60\degr < \ell < 360\degr$ (N = 157)
we got the same results as in Table~\ref{tab:2} to within the
1$\sigma$-errors, but the range in $\ovnec$ extends to $\ovnec
= 4\cmcube$, twice that of the corrected sample, and $\ovfilfac$ is
correspondingly lower. For our original, full sample of 285 pulsars we
find $\ovfilfac (\ovnec=1) = 0.0223\pm 0.0010$, exponent $= -1.02\pm
0.02$, $r= 0.94\pm 0.02$ and $t=49$. Even for the small sample of 13
distance calibrators, 5 of which are at $|b| < 5\degr$ (see
Table~\ref{tab:1}), we obtain a highly significant inverse correlation
in agreement with the larger samples within 3$\sigma$-errors. Thus also
without any corrections to EM a relationship between $\ovfilfac$
and $\ovnec$ with the correct slope can be derived for the DIG, but the
individual values of  $\ovfilfac$ and $\ovnec$ will be wrong by a
factor depending on the pulsar sample.

The inverse relationship between $\ovfilfac$ and $\ovnec$ derived here
refers to mainly interarm regions with relatively low average electron
density, $\avnel = 0.021\cmcube$. In spiral arms, where $\avnel$ seems
2--3 times higher (Cordes \& Lazio\ \cite{cordes+lazio02}),
correspondingly higher filling factors are expected if the slope of the
relation is near $-1$.

Which physical process or processes could cause the tight, inverse
relationship between $\ovfilfac$ and $\ovnec$? Without going into
details we mention two possibilities:

\begin{enumerate}
\item[---] Thermal pressure equilibrium in the DIG. If the temperature
increases, the clouds expand and the electron density in clouds
decreases. In Sect.~4.3.1 we gave some evidence for an increase of the
temperature with $|z|$.
\item[---] Turbulence in the ISM causing fractal structure. Much of the
ISM appears to have fractal structure which leads to $\ovfilfac
\propto \ovnec^{~-1}$ (Fleck\ \cite{fleck96}; Elmegreen\
\cite{elmegreen98}, \cite{elmegreen99}).
\end{enumerate}

Determinations of the $\ovfilfac (\ovnec )$-relation for other phases
of the ISM will be needed to see how widely it is valid. The physical
processes causing the inverse relationship $\ovfilfac \propto
\ovnec^{~-1}$ could be studied in simulations modelling the ISM.

\section{Summary and conclusions}
We have used dispersion measures of pulsars (Hobbs \& Manchester\
\cite{hobbs+manchester03}), distances from the model of Cordes \& Lazio
(\cite{cordes+lazio02}) and emission measures from the WHAM survey
(Haffner et al.\ \cite{haffner+03}) for a statistical study of several
properties of the diffuse ionized gas (DIG) in the Milky Way. The
emission measures were corrected for absorption and contributions from
beyond the pulsar as described in Sect.~2.3. To avoid regions with
excessive absorption and contributions from bright \ion{H}{ii} regions,
we selected pulsars at Galactic latitude $|b|>5\degr$. The final sample
analyzed contains 157 pulsars in the longitude range $60\degr < \ell <
360\degr$. Their distribution projected on the Galactic plane is shown
in Fig.~\ref{fig:2}. The statistical results are given in
Table~\ref{tab:2}.

Our main conclusions are summarized below.

\noindent
1.\  Dispersion measure DM and corrected emission measure
$\EM_\mathrm{p}$ are well correlated (Fig.~\ref{fig:5}b), indicating
that they probe the same ionized regions. This allows us to derive and
compare the average densities along the line of sight $\avnel$ and
$\langle\nel^2\rangle$, the mean electron density in clouds along the
line of sight $\ovnec$, the volume filling factor of these clouds
$\ovfilfac$ and the total path length through the ionized regions
$L_\mathrm{e}$ (see Eqs.~(1)--(4)).

\noindent
2.\  The total extent of the ionized regions perpendicular to the
Galactic plane increases from about 6~pc towards $|z| = 100\pc$ to about
220~pc towards $|z| = 1\kpc$, while the mean cloud density $\ovnec$
decreases from about $0.35\cmcube$ to about $0.10\cmcube$ (Sect.~3.4).

\noindent
3.\ Below $|z| \simeq 0.9\kpc$ $\avnel$ is
essentially constant with a maximum spread of a factor 2 about the mean
value of $0.021\pm 0.001\cmcube$ (Fig.~\ref{fig:10}). As $\avnel =
\ovfilfac \ovnec$,
this means that $\ovfilfac$ and $\ovnec$ are inversely correlated. While
$\ovnec$ decreases with increasing $|z|$ (Fig.~\ref{fig:7}b),
$\ovfilfac$ increases with $|z|$ (Fig.~\ref{fig:7}d). We derived the
relation $\ovfilfac (\ovnec ) = (0.0184\pm 0.0011)
\ovnec^{~-1.07\pm0.03}$, which is very tight. Figure~\ref{fig:8} shows
that it holds for the ranges $0.03 \la \ovnec < 2\cmcube$ and
$0.8 \ga \ovfilfac > 0.01$. The inverse dependence of $\ovfilfac$ on
$\ovnec$ could mean that the DIG is in thermal pressure equilibrium or
that it has a turbulent, fractal structure.

\noindent
4.\  The linear increase of $\DM\sin|b|$ with $|z|$ up to $|z|
\simeq 0.9\kpc$ indicates that the local electron density $\nel (z)$ is
constant, so also locally $\nec (z)$ and $f(z)$ are inversely
correlated. We suggest that the turn-over of $\DM\sin|b|$ near $|z| =
0.9\kpc$ occurs when $f(z)$ reaches a maximum value, whereas $\nec
(z)$ continues to decrease towards higher $|z|$ causing a flattening of
$\DM\sin|b|(z)$.

\noindent
5.\  Using this interpretation and the mean of the absorption-corrected
emission measures through the full layer $\overline{\EM_\mathrm{c}
\sin|b|}$, we derived a true scale height of $\nec (z)$ and $f(z)$
between $250\pc$ and $\simeq 500\pc$ for maximum values of $f(z)$
between 1 and 0.3, respectively, near $|z|\simeq 1\kpc$.
Since $\nel^2(z) = f(z) \nec^2 (z)$ and $\nel (z) = f(z) \nec (z)
=$~constant, $\nel^2(z)$ has the same scale height as $\nec (z)$
(Sect.~4.2). Beyond $|z|\simeq 1\kpc$, where $f(z) \simeq$~constant, the
scale height of $\nel^2 (z)$ becomes a factor of 2 smaller than that of
$\nec (z)$.

\noindent
6.\  The variation of $\ovfilfac (z)$ is similar to that of the
average degree of ionization of the warm, atomic gas,
$\oviw (z)$ (Fig.~\ref{fig:11}). Towards $|z| = 1\kpc$ $\ovfilfac =
0.24\pm 0.05$ and $\oviw = 0.24\pm 0.02$. The increase of
$\oviw$ with $|z|$ may point to an increase of the temperature in the
DIG with increasing distance to the Galactic plane. The local
function $I_\mathrm{w}(z)$ reaches values near unity at $|z| = 1\kpc$
(Table~\ref{tab:4}). If $f(z)$ reaches a maximum value when the
gas is nearly fully ionized, then the $|z|$-distance where this occurs
is indicated by the turn-over point of $\DM\sin|b|(z)$.

As the tight inverse relation between $\ovfilfac$ and $\ovnec$ is
independent of $|z|$ or distance along the plane, it seems a basic
property of the DIG. Simulations modelling the DIG may be able to
reproduce this relationship.

\begin{acknowledgements}
We thank Dr. A. Fletcher for useful discussions and careful reading of
the manuscript, and an anonymous referee for very extensive comments
leading to improvement of the paper. The Wisconsin H-Alpha Mapper is
funded by the National Science Foundation.
\end{acknowledgements}


\begin{appendix}

\onecolumn

\section{Glossary}

\begin{tabbing}
\=$\langle n_\mathrm{HI,w}\rangle$ \=\hspace{1cm}\= \kill
\> $A$   \> absorption correction applied to $\EM$\\
\> $B$   \> correction for emission measure from beyond the pulsar\\
\> $D$   \> distance to pulsar\\
\> $\DM$ \> observed dispersion measure\\
\> $\EM$ \> observed emission measure\\
\> $\EM_\mathrm{c}$ \> $\EM$ corrected for absorption, $\EM_\mathrm{c}
                       = \EM \cdot A$\\
\> $\EM_\mathrm{p}$ \> $\EM$ corrected for absorption and contributions
                        from beyond the pulsar, $\EM_\mathrm{p} = \EM
                        \cdot AB$\\
\> $f$         \> local volume filling factor\\
\> $\fdee$     \> fraction of line of sight containing electrons\\
\> $\ovfilfac$ \> average volume filling factor, average fraction
                  of radio beam up to pulsar containing electrons\\
\> $H$         \> exponential scale height of quantities averaged along
                  the line of sight\\
\> $h$         \> true scale height, exponential scale height of local
                  quantities\\
\> $I_\mathrm{w}$  \> local degree of ionization of warm \ion{H}{i}\\
\> $\oviw$     \> degree of ionization of warm \ion{H}{i} averaged along
                the line of sight\\
\> $L_\mathrm{e}$  \> total path length through ionized regions along $D$\\
\> $\nec$          \> local electron density in a cloud\\
\> $\ovnec$        \> mean electron density within clouds\\
\> $\nel$          \> local electron density, $\nel = f \nec$ and
                      $\nel^2 = f \nec^2$\\
\> $\avnel$    \> average electron density along the line of sight,
                  $\avnel = \ovfilfac \ovnec$\\
\> $\langle\nel^2\rangle$  \>average along the line of sight of the
                             square of the local electron density,
                             $\langle\nel^2\rangle = \ovfilfac \ovnec^2$\\
\> $n_\mathrm{HI,w}$     \>  local density of warm \ion{H}{i}\\
\> $\langle n_\mathrm{HI,w}\rangle$ \> density of warm \ion{H}{i}
                              averaged along the line of sight\\
\> $z$            \> height above the Galactic plane\\
\> $z_\mathrm{p}$ \>  height of the pulsar\\
\> $z_1$          \>  height at which $f$ reaches its maximum value\\
\end{tabbing}
\end{appendix}

\end{document}